\documentclass[a4 paper, 11pt,oneside]{article}
\usepackage[numbers]{natbib}
\usepackage{fullpage}  
\usepackage[lmargin=1.0in,rmargin=0.6in,tmargin=0.6in,headsep=.2in]{geometry}
\usepackage{graphicx}
\usepackage{forloop}
\usepackage{etoolbox}
\usepackage{calc}
\usepackage{wrapfig,lipsum,booktabs} 
\usepackage{xtab} 
\usepackage[dvipsnames]{xcolor}
\usepackage[normalem]{ulem}
\usepackage{sectsty}
\makeatletter
\def\@seccntformat#1{\@ifundefined{#1@cntformat}%
{\csname the#1\endcsname\;}
{\csname #1@cntformat\endcsname}
}
\def\section@cntformat{\thesection.\;} 
\def\subsection@cntformat{\thesubsection.\;} 
\makeatother
\usepackage{hyperref}
\hypersetup{
    colorlinks=true,
    urlcolor=blue,
    linkcolor=black,
    citecolor = black}

\usepackage{fancyhdr}
\fancypagestyle{first}{%
  \fancyhf{}
}
\usepackage{amsmath,amsfonts,amssymb,amsthm,epsfig,epstopdf,url,array}
 \usepackage[retainorgcmds]{IEEEtrantools}
 \usepackage{makeidx,epsfig,lscape}
 \usepackage{xcolor,pict2e}
\usepackage{upgreek}

\newcommand{\dd}{\textnormal{\,d}}
\newcommand{\erf}{\textnormal{\,erf}}
\newcommand{\hh}{\textnormal{H\textsubscript{I}}}
\theoremstyle{definition}

\begin{document}
\thispagestyle{first}
\vspace*{3cm}
{\noindent\huge\bf The Dynamics of Globular Clusters and Elliptical Galaxies}\\[1cm]
{\bf\large Marr John H.}\\[0.5cm]
Unit of Computational Science \\
Hundon, Suffolk, CO10 8HD, UK \\
Email: john.marr@2from.com \\

{\color{Black}\rule{0.7\textwidth}{2pt}}\\[0.2cm]
{\color{Black}\bf\large Abstract}\\
A model is developed for an idealised spherical galaxy evolving from a uniform mass distribution at the epoch of galactic separation until attaining an equilibrium state through gravitational collapse. 
The final theoretical radial surface density is computed and shows a good fit to the observational data for two globular clusters, M15 and M80. 
The mean cycle time and velocity are computed, the velocity-radius curve is developed and Gaussian RMS values derived, from which half-light radius vs. mass are plotted for 544 ellipticals plus compact, massive, and intermediate-mass objects. 
These show a linear mean log-log $R$-$M_{vir}$ slope of ${0.604\pm0.003}$, equivalent to a Faber-Jackson slope of $\gamma=3.66{\pm}0.009$ over a mass range of 7 decades. and a slope of $0.0045\pm0.0001$ on a semi-log plot of $R_{1/2}-\sigma$. 
Globular clusters, dwarf elliptical and dwarf spherical galaxies show a distinct anomaly on these plots, consistent with the ellipticals containing a supermassive black hole (SMBH) whose mass increases as the velocity dispersion increases, compared with the remaining types of spherical or irregular galaxies without a massive core. 
Analysis of the equations of motion suggested the generalised rule that all spherical galaxies expand from their initial radius at the epoch of galactic separation to a stable maximum radius of $\simeq1.136$ times their initial radius in their relaxed state.
\vspace{0.5cm}\\
{\color{Black}\bf\large Keywords}\\
Globular Clusters, Elliptical Galaxies, GC Formation, Virial Equation, GC Dynamics
\vspace{0cm}\\
{\color{Black}\rule{0.7\textwidth}{2pt}}

\section{Introduction}
\label{sec:Intro}
Globular clusters (GCs) are spherical collections of stars that orbit a galactic core as a satellite. 
They are fairly common structures, with over 150--157 currently known in the Milky Way (MW) \cite{2021A&A...649A...1G, 2023yCat..74894367P}. 
GCs are tightly bound by gravity, giving them their spherical shapes and relatively high stellar densities toward their centres.
They all have a similar appearance with spherical symmetry, a core concentration of stars, and a reasonably well defined edge to the star field.
They may contain a high density of stars: on average about 0.4 stars per cubic parsec, increasing to 100--1000 stars per cubic parsec in their core \cite{2023yCat..74894367P}, with a typical distance between stars of about one light year \cite{1942idlb.book..313C, 2020...Smail}. 
GCs may be categorised according to their degree of concentration toward the core, ranging from Class I for the most highly concentrated (e.g. M75), through successively diminishing concentrations to Class XII (e.g. Palomar 12).

Although the formation of GCs remains poorly understood, they contain some of the oldest stars in the Galaxy with a lack of O and B type stars and an abundance of low-mass red stars and intermediate-mass yellow stars, with an age estimated at $>10-13$ billion years \cite{1999idlb.book..832C, 2023arXiv230104659W}. 
Generally, GCs are free of gas and dust and do not display active star formation, being composed of old, metal-poor stars containing hydrogen and helium but not much else, and most of their stars are at approximately the same stage in stellar evolution suggesting that they formed at about the same time \cite{2001ASPC..245..162C}. 

These globular star clusters may contain $10^5-10^6$ members making detailed star-by-star simulations too complex for simple analysis, although approximate N-body simulations with several thousand particles have been performed \cite{2001ASPC..228.....D, 2014MNRAS.440.3172Z}. 
In practice, however, deviations from true thermodynamic equilibrium are small and N-body simulations show a relatively smooth behaviour of the approximately thermodynamic parameters as a function of time. 
The spherical Jeans Equation provides a good thermodynamic analysis for the motion of stars in 3-dimensions, but a specific solution is difficult \cite{2008gady.book.....B}; for this reason, various approximation methods, such as Fokker-Planck treatments and conducting gas sphere models, have worked well \cite{1997Comp....3.1H}. 
By making some justified assumptions for the initial density distribution and kinetic parameters of an idealised GC, we build an analytic model that is consistent with observations for a number of GCs and comparable to the Jeans Equation.

\subsection{Chronology of GCs}
\label{Chronology}
Globular clusters are old structures, with ages comparable to the absolute age of the universe ($t_U$), a cosmological limit setting a maximum age to all objects within it and inferred by Planck-18 data to be $t_U = 13.800\pm0.024$ Gyr \citep{2020A&A...641A...6P}.
The terminal phase of the early Universe was dominated by high energy $\gamma$-photons coupled to free protons and electrons (the photon epoch) until, with continuing expansion and cooling, the remaining free protons and electrons transitioned to neutral atomic ($\hh$) and molecular ($H_2$) hydrogen at the epoch of photon decoupling, leading to the era of recombination at redshift $z=1091.64\pm0.47$, $\sim375,000$ years after the Big Bang \citep{2013ApJS..208...20B}.
Photons then ceased to significantly interact with matter allowing the Cosmic Microwave Background (CMB) radiation to penetrate without further major interaction, with an extremely isotropic and homogeneous thermal black body spectrum at its modern value of about $2.73^\circ$K, strongly supporting the CMB to be a remnant of the Big Bang. 
This sets a further age limit because, apart from the small ripples observed in the microwave background radiation (MWB), the whole Universe was then in a state of equilibrium \citep{2020A&A...641A...6P}. 
This is uncomfortably close to the ages of some observed components of the Universe, such as some supermassive black hole (SMBH) quasars, some of the oldest stars, and some of the oldest GCs, leaving a very short time for them to have condensed from the expanding, hot gas field.

Finer details of the anisotropies in the background radiation have become apparent with the increasing sensitivity of COBE, the Wilkinson Microwave Anisotropy Probe (WMAP) satellite project, and the Planck cosmology probe \cite{2009ApJS..180..225H, 2016A&A...594A..13P} giving hints of early structure and consequent galactic separation, possibly through the amplification of quantum fluctuations within the proto-universe itself.
Reionisation occurred relatively quickly as objects began to condense in the early universe and became sufficiently energetic to re-ionise neutral hydrogen, with recent results from the Planck mission yielding an instantaneous reionisation redshift of $z = 7.68\pm0.79$ \cite{2018arXiv180706209P}. 

It is generally accepted that the expanding Universe differentiated into gas clouds that subsequently collapsed, but the formation of quasars of mass $>10^{10}$ M$_\odot$ and redshifts $>7$ would require an accretion time greater than their age at that epoch.
Finkelstein {\it{et al}} identify a sample of 26 galaxy candidates at $z\sim9-16$ \cite{2022arXiv221105792F}, and one of the oldest galaxies with an age of 13.39~Gys ($z=11.09^{+0.08}_{-0.12}$) has been identified by a continuum Ly$\alpha$ break just $\sim$400~Myr after the Big Bang, suggesting that galaxy build-up was well underway early in the reionization epoch \cite{2016ApJ...819..129O}.
Quasars are extremely luminous active galactic nuclei (AGN), with a supermassive black hole (SMBH) ranging from millions to billions of times the mass of the Sun.
The most distant quasar currently known is J0313-1806 at $z=7.6423\pm0.0013$ when the universe was only $\sim670$~Myrs old \cite{2021ApJ...907L...1W}, and it is likely that these SMBH objects were the first to condense due to their high mass.
The origin of the SMBHs that power distant quasars remains uncertain, and their rapid formation presents a paradox because the ages of the oldest quasars, stars and GCs extends into the period before the end of reionisation \cite{2019PASA...36...27W}, although some recent work has suggested a solution to the problem of early quasar appearance using a classical approach \cite{2020MNRAS.498.5652K}.

The earliest population III (Pop III) massive stars cannot be older than $t_U$, and are believed to have formed in dark matter mini-halos of typical mass $M=10^5-10^6$M$_\odot$  at redshifts $z=20-30$, $\sim100$ million years after the Big Bang at the end of the cosmic dark ages.
New parallaxes from the Gaia space mission that replace the older HST measurements assess the age of the Milky Way halo Population II, metal deficient, high velocity subgiant HD-140283 (or Methuselah star) to be $12.0\pm 0.5$ Gyr, or $1.8\pm0.5$~Gyr after the Big Bang \citep{Tang_2021}, comfortably within $t_U$ and after the CMB.
Analysis of very low-metallicity stars in globular clusters suggests their age to be $13.32\pm0.1$~Gyr, with $t_U$ constrained to be $13.5^{+0.16}_{-0.14}$ Gyr \citep{2020JCAP...12..002V}. 

Nevertheless these ages are all uncomfortably close to the epoch of decoupling, suggesting that stars of the earliest GCs formed rapidly from the condensing gas clouds against the expanding universe.
This early formation timescale for old GCs and ancient stars, galaxies, and quasars has suggested some new mechanisms such as supra-exponential accretion with fissuring of the hot plasma itself as the Universe expanded, rather than re-condensation of the expanding hot gas \cite{2014Sci...345.1330A}.

\subsection{Scarcity of massive stars in Globular Clusters}
The possible existence of a black hole at the centre of GCs has been considered by a number of theoreticians \cite{Gerssen_2002}.
Although some theoretical models suggested that GCs do not require a black hole core, a recent spectroscopic search of 25 Galactic GCs discovered a binary system in NGC~3201, one component of which is thought to be a detached stellar-mass black hole with a minimum mass of $4.36\pm0.41$~M$_\odot$ \cite{10.1093/mnrasl/slx203}.

A number of mechanisms account for the paucity of massive stars in GCs: 
(1)~the initial number of massive stars (i.e.~$>10$~M$_\odot$) was probably low as a statistical consequence of the GCs being small systems with few stars; 
(2)~the age of GCs implies that no early stars with initial mass $>2.5$~M$_\odot$ will still be present in a small GC, as they will have formed supernovae shortly after the formation of the GC; 
(3)~the low mass of GCs may have prevented the retention of secondary elements formed by early supernovae and hence any consequent population~I secondary star formation will have been suppressed; 
(4)~conversely, stars with very low mass ($<\approx 0.1$~M$_\odot$) may not have ignited, and were possibly sling-shot out from their parent GC. 
Recent observations with the JWST show that M92, for instance, with a mass resolution of $0.002M_\odot$, shows a minimum mass for hydrogen burning of $0.078 \leq M\leq0.08M_\odot$, a metallicity of [Fe/H]= -1.7 dex and an age of 13~Gy \cite{2023arXiv230104659W}.

The cross sectional area of a star of stellar mass is $1.5\times 10^{18}$~m$^2$. 
Collisions between stars in a GC are probably rare, as confirmed by the very low rate of super novae \cite{2012A&A...539A..77V}.
For M15 with a total stellar mass of $4.5\times 10^5$~M$_\odot$ and radius of 25~pc, this suggests that the mean distance between stars in the core of even a large GC is $\sim1.5\times 10^{12}$~m, with a mean cross sectional area between stars of $\sim 1.32\times 10^{36}$~m$^2$.
The ratio of areas between this and the star is then $\sim 8.8\times10^{17}$, implying a low probability for accretion of core stars onto a black hole, therefore the binary black hole in NGC~3201 may have a similar age to its component stars.

\section{Assumptions of the model}
\label{sec:assumptions}
The virial theorem is a powerful tool for analysing the motion of stars in an idealised galaxy, where the potential energy ($PE$) is generally taken to be the $PE$ of a mass $m$ brought from infinity towards a self-gravitating galaxy of total mass $M$ and this defines the zero point, $U$, for $PE$.
Assuming that all mass within a sphere defined by $r$ is equivalent to a point mass at the galactic centre and all mass external to that sphere may be ignored, the usual gravitational force $F\propto r^{-2}$ may be applied and the standard equations of motion for the star system derived \cite{2015MNRAS.448.3229M}.

In the early universe however, defined here as the epoch of galaxy separation, ($T_g$), this standard analysis for $U$ may not be applicable because, at this epoch of separation, the density of the universe was approximately uniform, with the exception of local fluctuations. 
We consider the expanding Universe to have separated relatively quickly into islands of matter round areas where the density was a little above average, with nucleation occurring as a precursor to the massive core at galactic centres; where the density was a little lower than average, the continuing expansion of the universe may be thought of as ripping through the distribution of matter as though the protogalaxies were being torn apart from each other due to the expansion of space at the time $T_g$.
We therefore consider the potential energy of any small mass, $m$, at a radial distance $r$ from the centre, to then be fixed in place during the expansion of the universe until galactic separation was sufficient for the islands to become self-gravitating at the era of separation. 
This model therefore modifies the definition of $U$ by assuming that individual masses ($m$) within each galaxy did not come from infinity, as the mean gravitational field was everywhere zero at this moment; but as neighbouring galaxies had reducing mutual gravitational influence, the initial $U$ for each $m$ would have continued to increase asymptotically towards its theoretical value had $m$ been brought in from infinity, thus driving gravitational collapse.

At $T_g$, it may be debated whether the matter of the universe was in the form of gas or if early star formation had already occurred.
The analysis for this model is not contingent on the time of star formation, and the model therefore assumes that these protogalaxies were populated by young, uniform stars, each of mass $m$. 
The uniform density of the universe at the epoch of galaxy separation, $T_g$, is defined as $\rho_0$, with the mean radius of each protogalaxy, $r_0$, containing a mass $M_0$. 
Although the individual masses had been separating from each other as a consequence of the overall expansion, at the epoch of galaxy separation it is assumed that the individual masses within each protogalaxy would cease expanding away from each other, and the mean kinetic energy within each galaxy at this instant before gravitational contraction was therefore zero. 
Thus the total mechanical energy of the system was derived ultimately from the separation of mass by the initial expansion of space, and the potential energy of these masses increased to its final value as neighbouring galaxies drew away. 
As separation continued, the masses began to infall through mutual self-gravitational collapse into the local gravitational field generated by the protogalaxy. 

This model assumes that self-gravitational infalling would continue, with the newly acquired initial potential energy converted into kinetic energy as the masses accelerated inwards. 
The motion of each individual star is highly chaotic and equipartition of $U$ throughout the system is assumed through gravitational interactions, such that the masses reach dynamic stability, until finally moving with a mean radial velocity, $\overline{v}$. 
At this time, it is shown that a state of equilibrium would be reached in which the stars would undergo periodic motion, traversing a path across a sphere of final radius $r_f$.
The model makes the following assumptions:
\begin{itemize}
  \setlength{\itemsep}{1pt}
  \setlength{\parskip}{0pt}
  \setlength{\parsep}{0pt}
\item Radial symmetry about the centre.
\item  No angular momentum.
\item The equations of motion depend wholly on Newtonian gravitation in standard form.
\item The galaxy starts at time $T_g$ (the epoch of galactic separation) from a collapsing, self-gravitational sphere of radius $r_0$.
\item The sphere collapses to a final equilibrium radius, $r_f$.  
\item The initial state of the protogalactic sphere may have been in the form of gas, but as this collapsed in on itself, star systems spontaneously formed which were each approximately the age of the GC. 
\item The whole system may therefore be treated as though it were initially composed of $n$ stars of mean mass $m$, with an initially uniform distribution of mass with initial density $\rho_0$.
\item Equipartition of energy such that, over sufficient time (but a time much shorter than the observed age of the galaxy), the total energy of the system is partitioned equally within each spherical shell, as a function of radius.
\item Conservation of mass: The mass of the galaxy remains $M_0=n\times{}m$ throughout. 
\item Conservation of energy: the loss in potential energy in collapsing from $r_0$ to $r_f$ is converted entirely into kinetic energy of motion of the stars.
\end{itemize}

\subsection{Equation of motion in a uniform spherical galaxy}
We assume the GC density (stars+gas) was approximately uniform at $T_g$, the epoch of galactic separation. 
The mean gravitational field within an infinite uniform distribution of mass is zero, but after $T_g$ each island galaxy became increasingly isolated, with a defined centre. 
Thereafter, a typical galaxy became a self-gravitating sphere of particles, such as a star field, with initial radius $r_0$ and initial uniform mean mass density $\rho_0$, which was the density of the Universe at the era of galactic separation.

\subsection{Initial potential energy of a spherically symmetrical galaxy}
The gravitational force on a test mass $m$ is:
\begin{equation}
    F=G\frac{mM_r}{r^2}\,,
	\label{eq:F}
\end{equation}
where $M_r$ is the mass internal to $r$:
\begin{equation}
    M_r=\frac{4}{3}\pi{}r^3\rho_0\,.
	\label{eq:Mr}
\end{equation}
 
The potential energy ($U$) of the mass $m$ is defined as the work done in bringing the mass from infinity to $r$, and is the radial path integral of $F$:	
\begin{equation}
	U=-G\frac{mM_r}{r}\,.
	\label{eq:PE}
\end{equation}
Hence, for a thin shell at $r$ of mass $m_{shell}$ and thickness $\delta{}r$, $\delta U$ will be:
\begin{equation}
	\delta{}U=-G\frac{M_r m_{shell}}{r}\,,
	\label{eq:dPE}
\end{equation}
\begin{equation}
	\textnormal{with a shell mass: }m_{shell}=4\pi\rho_0r^2\delta{}r\,.
	\label{eq:5}
\end{equation}
\begin{equation}
	\textnormal{Hence: }\delta{}U=-G\frac{16\pi^2\rho_0^2r^4}{3}\delta r\,,
	\label{eq:6}
\end{equation}
and the total initial $U$ is: 
\begin{equation}
	U_0=-G\int_0^{r_0}\frac{16\pi^2\rho_0^2r^4}{3}\dd r
	=-G\frac{16\pi^2\rho_0^2r_0^5}{15}\,.
	\label{eq:7}
\end{equation}
But the total mass of the galaxy is $M_0=\frac{4}{3}\pi{}r_0^3\rho_0$,
\begin{equation}
	\textnormal{i.e. }\rho_0=\frac{3M_0}{4\pi{}r_0^3}\,,
	\label{eq:9}
\end{equation}
\begin{equation}
	\textnormal{Hence: }U_0=-G\frac{3M_0^2}{5r_0}\,.
	\label{eq:10}
\end{equation}

\subsection{Final potential energy}
\begin{figure}
	\includegraphics[width=0.5\columnwidth]{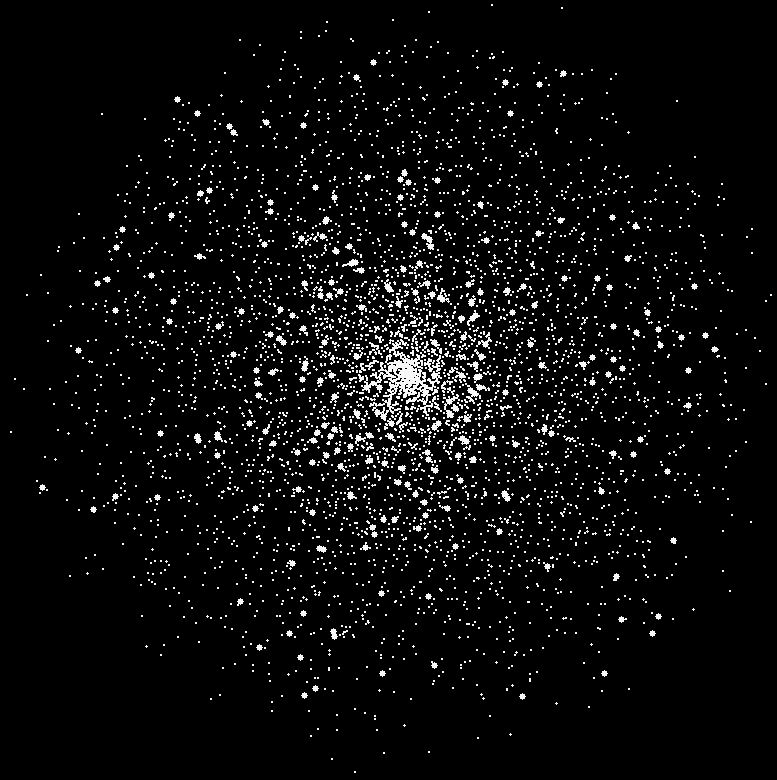}
    \vspace{0cm}
    \centering
    \caption{Monte Carlo simulation of a globular cluster of 5000 stars with shell density $\propto r^{-2}$ and simple IMF distribution without mass segregation}
    \label{fig:1}
\end{figure}

Making the assumption that GCs began as spheres of uniform density, these would have been dynamically unstable leading to collapse of the sphere. 
At the initial uniform density each shell would have contained a mass of stars $m_{shell}\propto{}r^2$, but at equilibrium the flux of stars moving into and out of a shell must exactly balance. 
This implies that the number density of stars in each shell should be $\rho_{shell}\propto r^{-2}$ as visualised in a 
Monte Carlo simulation using this $\rho \propto r^{-2}$ distribution for 5000 stars with a simple Salpeter IMF (Fig.~\ref{fig:1}), which is similar to the distribution of visible stars in a classical globular cluster such as M92.

Using the equilibrium distribution of $\rho \propto r^{-2}$, we define the density function:
\begin{equation}
	\rho(r)=\rho_fr_f^2/r^2\,,
	\label{eq:density}
\end{equation}
where $\rho_f$ is a characteristic final density at the surface of the galaxy, at final radius $r_f$. 
The total mass is then:
\begin{equation}
	M_{tot}=\int_0^{r_f}4\pi{}r^2\rho(r)\dd r=M_0\,,
	\label{eq:12}
\end{equation}
\begin{equation}
	\textnormal{i.e. }M_{tot}= 4\pi\rho_fr_f^3=M_0=\frac{4}{3}\pi\rho_0r_0^3\,,
	\label{eq:13}
\end{equation}
\begin{equation}
	\textnormal{and hence: }\rho_f=\frac{1}{3}\rho_0\left(\frac{r_0}{r_f}\right)^3=\frac{M_0}{4\pi{}r_f^3}\,,
	\label{eq:14}
\end{equation}
where $r_0$ and $\rho_0$ are the initial radius and initial mean mass density respectively.
Substituting
\begin{equation}
	M_r=\int_0^r 4\pi{}r^2\rho(r)\dd r =4\pi\rho_f r_f^2 r
	\label{eq:16}
\end{equation}
\begin{equation}
	\textnormal{and~~~~~}m_{shell}=4\pi\rho(r) r^2\delta{}r=4\pi\rho_f r_f^2\delta{}r
	\label{eq:17}
\end{equation}
into Eq.~(\ref{eq:dPE}) for the potential energy of a shell at $r$, thickness $\delta{}r$, then:
\begin{equation}
	\delta{}U=-16\pi^2G\rho_f^2 r_f^4\delta{}r\,.
	\label{eq:18}
\end{equation}
The final total potential energy is therefore:
\begin{equation}
	U_f=-16\pi^2G\rho_f^2 r_f^5 =-GM_0^2/r_f\,,
	\label{eq:19}
\end{equation}
and the change of $U$ is:
\begin{equation}
	\Delta{}U=-GM_0^2\left(\frac{3}{5r_0}-\frac{1}{r_f}\right)\,.
	\label{eq:20}
\end{equation}

\subsection{Kinetic energy}
The total kinetic energy for $n$ stars of mean mass $m$ at equilibrium is:
\begin{equation}
	K_{tot}=\sum_1^n \frac{1}{2} m v_r^2
	\label{eq:21}
\end{equation}
Assuming full equipartition of energy amongst the stars in each shell, we may consider a mean velocity $\overline{v}$ at equilibrium such that:
\begin{equation}
	K_f=\frac{1}{2}M_0 \overline{v}^2
	\label{eq22}
\end{equation}
\begin{equation}
    \textnormal{where }\overline{v}^2=\frac{1}{n}\sum{v_r^2}\,.
\end{equation}
Equating total $K$ with the decrease in $U$:
\begin{equation}
	\overline{v}^2=\frac{2K_f}{M_0}=2GM_0\left(\frac{1}{r_f}-\frac{3}{5r_0}\right)
	\label{eq:v_bar1}
\end{equation}

\subsection{Periodic motion of stars within a sphere with increasing density}
The equation of motion of a star of mass $m$ within a gravitational sphere of density $\propto r^{-2}$ is given by:
\begin{equation}
	F=m\ddot{r} =G\frac{mM_r}{r^2} = \frac{4\pi{Gm}\rho_f r_f^2}{r}
	\label{eq:24}
\end{equation}
\begin{equation}
	\textnormal{or }\ddot{r}=\frac{a_0}{r}\,,
	\label{eq:25}
\end{equation}
where $a_0=4\pi{}G\rho_f r_f^2$.
This may be solved analytically by considering the change in $U$ of $m$ as it falls across the sphere, which is the work done in bringing the test mass $m$ towards the centre against the gravitational field, $F$. 
Defining the zero of $U$ to be at the galactic surface, $r_f$, and $\Delta{}U$ to be the change in $U$ in moving from $r_f$  to $r$, then,
\begin{equation}
	\Delta{}U=m\int_{r_f}^r\frac{a_0}{r} \dd r\,,
	\label{eq:26}
\end{equation}
\begin{equation}
	\textnormal{i.e. }\Delta{}U=-ma_0\ln\left(\frac{r_f}{r}\right)\,.
	\label{eq:27}
\end{equation}
Using the conservation of energy, $\Delta{}K + \Delta{}U=0$,
\begin{equation}
	\Delta{}K=m a_0 \ln(r_f/r)= \frac{1}{2} mv^2\,.
	\label{eq:28}
\end{equation}
Rearranging, and using $v=dr/dt$,
\begin{equation}
	\dd t=\frac{\dd r}{\sqrt{2a_0 \ln(r_f/r)}}\,,
	\label{eq:29}
\end{equation}
which has a standard integral solution,
\begin{equation}
	\int{}\frac{\dd r}{\sqrt{\ln(a/r)}}=-a\sqrt{\pi} \erf{\left(\sqrt{\ln(a/r)}\right)}\,.
	\label{eq:30}
\end{equation}
The $\erf{(z)}$ is the integral of the Gaussian distribution, i.e.,
\begin{equation}
	t=\left|-r_f\sqrt{\frac{\pi}{2a_0}} \erf{\left(\sqrt{\ln{(r_f/r)}}\right)}
    \right|_{r=r_f}^{r=0}
	\label{eq:31}
\end{equation}
which is a continuous and complete function.
Noting that $\erf{(0)}=0$, and $\erf{(\infty)}=1$, 
the time for $m$ to fall towards the centre (a quarter cycle) is therefore,
\begin{equation}
	T_{1/4}=\pm\sqrt{\frac{1}{8G\rho_f}}
	\label{eq:32}
\end{equation}
where $T$ is the time for a full cycle. 
The emergence of the error function in this integral might be anticipated from remembering that the velocity of any particle in a system at thermal equilibrium will have a normal distribution due to the maximum entropy principle. 

The mean velocity is then $r_f/T_{1/4}$, and the mean square velocity is:
\begin{equation}
	\overline{v}^2=8G\rho_f r_f^2=\frac{2GM_0}{\pi{}r_f}\,.
	\label{eq:v_bar}
\end{equation}
But from Eq.~(\ref{eq:v_bar1}):
\begin{equation}
    \overline{v}^2=2GM_0 \left(\frac{1}{r_f}-\frac{3}{5r_0}\right)\,,
	\label{eq:34}
\end{equation}
\begin{equation}
	\textnormal{i.e. } \frac{1}{\pi{}r_f}=\left(\frac{1}{r_f}-\frac{3}{5r_0}\right)\,. 
	\label{eq:35}
\end{equation}
Substituting $\epsilon=r_f/r_0$  and solving for $\epsilon$:
\begin{equation}
	\epsilon=\frac{5}{3}\left(1-\frac{1}{\pi}\right)\approx1.136
	\label{eq:epsilon}
\end{equation}
This is a remarkable result, implying that all simple spherical galaxies expand to have a stable maximum radius of $\approx1.136$ times their initial radius at the epoch of galactic separation. 
Although physically larger, the majority of the mass is internal to the initial radius and the total potential energy has decreased, while equipartition of kinetic energy ensures that this concentration of mass is built by more massive stars congregating towards the centre of the GC \cite{2008gady.book.....B}.

Although this simplistic approach ignores relativistic effects as the mass accelerates towards the centre, in practice the motion is highly chaotic through the action of gravitational interactions and from any non-radial component inherited from the time of galaxy formation. 
This therefore causes the star to swing past the centre, apart from very rare encounters with the event horizon at a Schwarzschild radius if a black hole core is present. 
There is therefore a low probability of any individual star experiencing relativistic encounters. 

\subsection{Time-distance and velocity-radius curves for M15}
\begin{figure}
\centering
    \includegraphics[width=0.5\columnwidth]{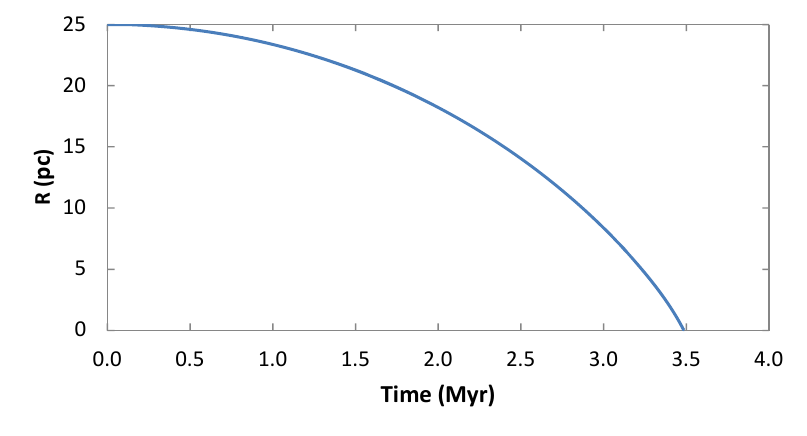}
    \vspace{0cm}
    \caption{Distance/Time curve for theoretical particles in M15 over quarter cycle $T_{1/4}$}
    \label{fig:RT}
\end{figure}
\begin{figure}
\centering
    \includegraphics[width=0.5\columnwidth]{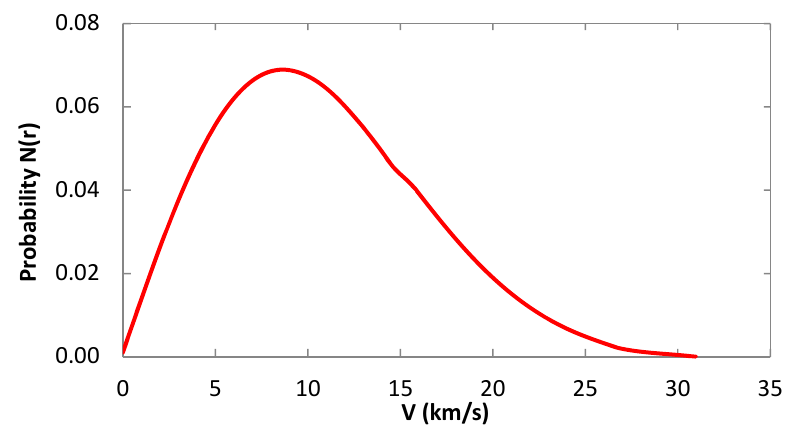}
    \vspace{0cm}
    \caption{Velocity-Probablility curve for M15}
    \label{fig:VP}
\end{figure}

Figure~\ref{fig:RT} shows the mean time-distance curve from Eq.~(\ref{eq:31}) for a typical star in galaxy M15, assuming a radius of 25~pc and a total mass of $4.5\times 10^5 M_\odot$.
The probability of finding a star with any given velocity increment is shown in Fig.~\ref{fig:VP}, normalised so that the area under the curve is 1 (i.e. certainty). 
This theoretical LoS probability curve across the centre of the galaxy is broadly in agreement with the observational velocity probability curve presented by  McNamara {\it{et al}} \cite{McNamara_2004}, and has similarities to a Rayleigh distribution, suggesting the distribution has a maximum Cumulative Residual Entropy \cite{2012JStatRes...B}. 
 More realistic curves to match observational values would introduce a mass distribution function for the stars but, although more complex, the fundamental dynamics would not change. 

\subsection{The IMF and globular clusters}
The Initial Mass Function (IMF) for the stellar component of early galaxies has been debated for more than 50~years since Salpeter proposed his initial model function: $N(dM)\propto M^\alpha$, where $N(dM)$ is the number of stars in the mass range $M$ to $M+dM~pc^{-3}$ \cite{1955ApJ...121..161S, 2005ASSL..327...41C}.
Detailed analysis of the different models has been provided in several papers\cite{2000A&A...363..476B, 2003PASP..115..763C, 2013pss5.book..115K}, but the model of  Miller {\it{et al}} is still commonly used with a 4-part segmentation of $\alpha$ for $M$ ranging from $0.1-100\,M_\odot$ \cite{1979ApJS...41..513M}.
Because of the rapid decline in number counts of more massive stars in this model, 52\% of the mass resides in stars with a mass $\leq 1 M_\odot$.

Applying an IMF function to the mass distribution of the dynamic model of Section 1, with equipartition of energy among the mass components, will change the characteristics of the density, radius and velocity profiles. 
There is an increase in core density and a more rapid reduction in shell density with radius; the core velocity is decreased slightly, whilst the final galactic radius is increased by the faster moving low mass components. 
However, although the exact profiles are IMF-dependent \cite{2013pss5.book..115K}, the overall picture remains well described by the simpler analysis using a single value for the mean stellar mass.

\section{Cross-sectional density}
Emission profiles of galaxies map surface luminosity per unit area as a function of galactic radius. 
These profiles are dependent on the volume density, but also on the mass-luminosity function and obscuration. 
With the assumption of uniform individual brightness profiles and no obscuration of deeper stars by dust, the number of stars per unit area as a function of radius in a galaxy of uniform density $\rho_f$, radius $r_f$ is simply:
\begin{equation}
    N(r)=2\rho_f \sqrt{r_f^2-r^2}\,,
    \label{eq:N_r3}
\end{equation}
where $N(r)$ is the number density of stars observed per unit area.

King\cite{1962AJ.....67..471K} presented an empirical formula for the density from centre to edge in GCs of the form:
\begin{equation}
    N(r)=k\left[\frac{1}{\sqrt{1+(r/r_c)^2}}-\frac{1}{\sqrt{1+(r_f/r_c)^2}}\right]^2
    \label{eq:King}
\end{equation}
where $N(r)$ is the number of star counts per unit area, $k$ is a constant, $r_c$ is a scale factor that may be called the core radius, and $r_t$ is the value of $r$ at which the surface density $f\rightarrow0$. 
With these three parameters, a good correlation was demonstrated with the observed number/radius counts for several globular clusters.
\begin{figure}
\centering
    \includegraphics[width=0.5\columnwidth]{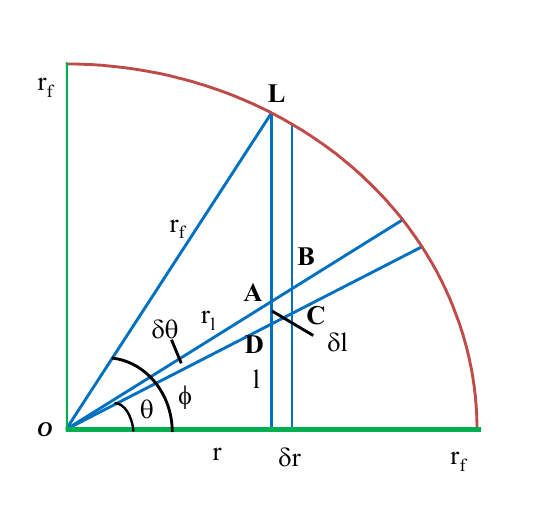}
    \vspace{0cm}
    \caption{Cross sectional star density in spherical globular cluster with variable density}
    \label{fig:fig5}
\end{figure}

Referring to Fig.~\ref{fig:fig5}, the number density of stars observed, $N(r)$, across a chord of the GC in a core of depth $2L$, unit area $\Delta A$, at a distance $r$ from the centre is:
\begin{equation}
    N(r)=2\int_0^L \rho(r_l)\delta l
    \label{eq:N_r2}
\end{equation}
where $\rho(r_l)$ is the density of the star field at radius $r_l=OA$ from the centre.
\begin{figure}
\centering
    \includegraphics[width=0.4\columnwidth]{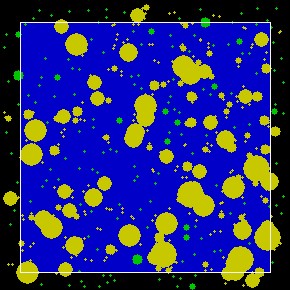}
    \caption{Simulated dense star field for star count corrections. Overlapping star fields in yellow, unobscured stars in green. Plate area $2\times2$ arc min$^2$}
    \label{fig:SF}
\end{figure}
Using the density relationship of Eq.~(\ref{eq:density}) and substituting: 
\begin{equation*}
\begin{split}
    L&=\sqrt{r_f^2-r^2}\,; \\
    \rho_{rl}&=\rho_f(r_f^2/r_l^2)\,; \\
    r_l^2&=r^2+l^2\,,
\end{split}
\end{equation*}
\begin{equation}
\begin{split}
    \textnormal{then     }N(r)&=2\rho_f \int_0^L\frac{r_f^2}{r^2+l^2} \dd l\\
    &=2\rho_f \frac{r_f^2}{r}\left|\tan^{-1}(l/r)\right|_0^L\,, 
\end{split}
\end{equation}
\begin{equation}
    \textnormal{i.e.~~~}N(r)=2\rho_f \frac{r_f^2}{r}\tan^{-1}\left(\frac{r_f^2}{r^2}-1\right)^{1/2}\,.
    \label{eq:41}
\end{equation}

Note that $N(r)\rightarrow0$ for $r\rightarrow r_f$, and for the limit $r\ll r_f$ this simplifies to:
\begin{equation}
    N(r)\approx\frac{\pi \rho_f r_f^2}{r}\,,
\end{equation}
which is a simple power curve with log-log slope $-1$. 
The implied limit $N(r)\rightarrow\infty$ as $r\rightarrow 0$ is not a physical reality; in practice, observed stars have a finite resolvable disk whose diameter is set by the resolving power of the telescope. 
In addition, it may be assumed that the star number density cannot increase beyond a certain limiting threshold, beyond which individual stars may conjoin, possibly as a central black hole.  
This sets an upper bound to the number of stars that can be distinguished in any given star field. 
If we define this radius to be $r_c$ , this represents a lower bound for $r$ in equation~\ref{eq:41}, from which we may generate the limiting curve:
\begin{equation}
    N(r)=2\rho_f  \frac{r_f^2}{(r+r_c)}\tan^{-1} \left(\frac{r_f^2}{(r+r_c)^2}-1\right)^{1/2}\,.
    \label{eq:43}
\end{equation}

The curve has three parameters, $\rho_f$, $r_f$, and $r_c$, and superficially resembles that of King but differs in concept.
King's curves were constructed {\it ad hoc} to fit the observational data, whereas the curve described in equation~\ref{eq:43} was generated from a model of the behaviour of stars in a relaxed spherical galaxy, and its shape differs in detail.

\subsection{Limitations of the model}
\begin{figure*}
\centering
    \includegraphics[width=12cm]{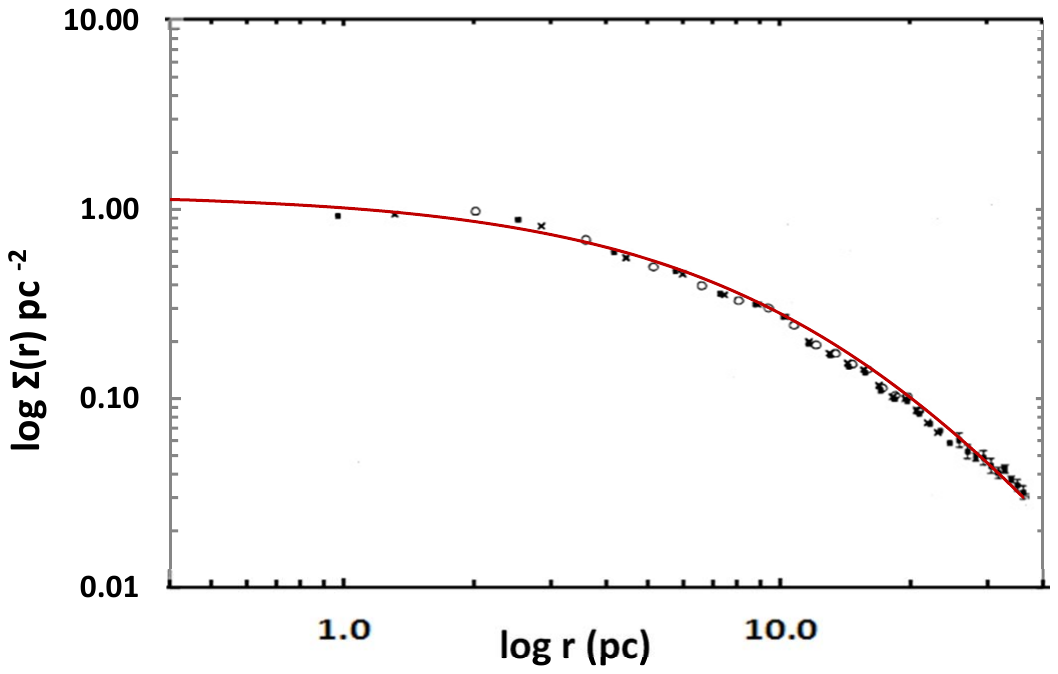}
    \vspace{0cm}
    \caption{Log-log surface density/radius for NGC6093 (M80) overlain with derived theoretical curve (equation~\ref{eq:N_r}). Different symbols referred to different plates in the original data. (Data from  \cite{1984ApJ...277L..49D})}
    \label{fig:NGC6093}
\end{figure*}
\begin{figure*}
\centering
    \includegraphics[width=12cm]{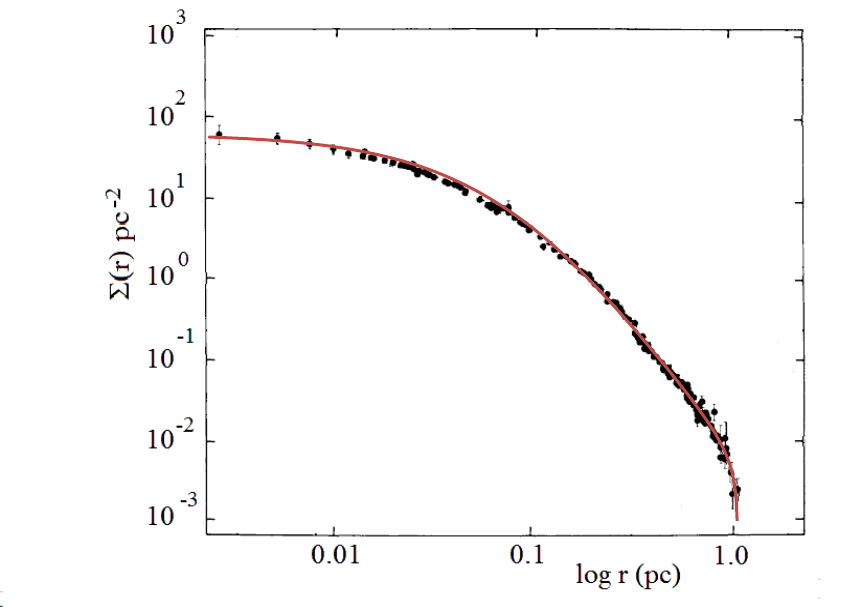}
    \vspace{0cm}
    \caption{Log-log surface density/radius for M15 overlain with derived theoretical curve (equation~\ref{eq:N_r}). Data adapted from   \cite{1976ApJ...208L..55N}}
    \label{fig:M15}
\end{figure*}
\begin{figure*}
\centering
    \includegraphics[width=12cm]{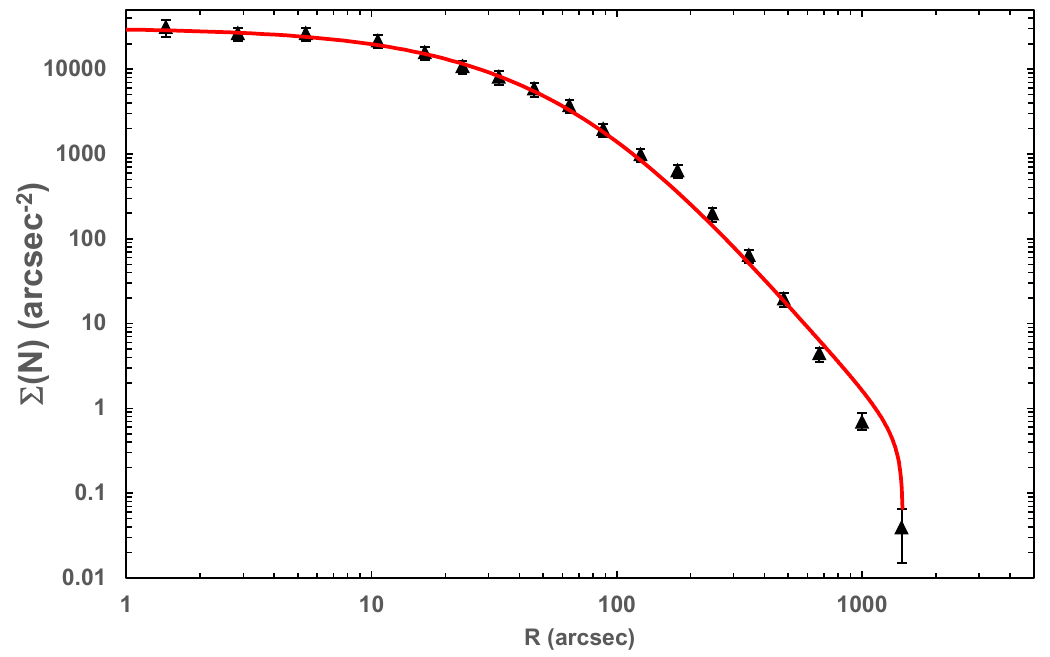}
    \vspace{0cm}
    \caption{Log-log surface density/radius for M92 with derived theoretical curve. Data adapted from \cite{Di_Cecco_2013}}
    \label{fig:M92}
\end{figure*}
One practical limitation for accurate number counts of stars in a crowded field comes from overlapping of the optical disks.
Figure~\ref{fig:SF} represents the simulated field of view of a plate $2\times2$ arc min$^2$, with a true star density of 100 stars arc min$^{-2}$. 
Unobscured stars are shown in green; the yellow stars represent an overlap of the optical disks, with the degree of overlap ranging from brushing edges to complete masking and this clearly affects the final counts. 
The stars outside the plate area are included in the figure to allow for completeness in the algorithm for assessing visible star counts. 
A majority of star disks show some degree of overlap but, by pixel counting in the plate area, an estimate may be drawn for the degree of overlap for any given star density and distribution of apparent radii. 

In common with King's model equations, equation~\ref{eq:43} assumes a definite maximum radius, $r_f$, a limiting edge density $\rho_f$, and a defined core radius, $r_c$. 
Implicit also is the assumption that kinetic energy is partitioned equally within each shell, and that all stars have identical masses. 
This latter is the least plausible assumption, and a better model can be created using an idealised initial mass function (IMF) such as that of Salpeter, with an empirical IMF $P(n)\propto m^{-2.35}\propto r^{-7.05}$, where $P(n)$ is the probability of observing $n$ stars of given mass $m$. 

The formal analysis of the equations of motion for stars of differing masses within globular clusters is complex, but by using an IMF of simplified Salpeter form such as $P(n)\propto m^{-2}$, a modified function for the predicted number counts per unit area may be derived (equation~\ref{eq:N_r}): 
\begin{equation}
    N(r)=\rho_f\frac{r_f^4+2r_f^2(r+r_c)^2}{3(r+r_c)^4}\sqrt{r_f^2-(r+r_c)^2}\,.
    \label{eq:N_r}
\end{equation}

Figs. \ref{fig:NGC6093}, \ref{fig:M15} and \ref{fig:M92} show three plots for the GCs NGC6093, M15 and M92 using this curve, with parameters adjusted for best fit to observational data. 

\subsection{Ratio of mean galactic separation to mean galactic radius}
At the era of separation, $T_g$, proto-galaxies had expanded to a radius $r_0$. 
$T_g$ corresponds approximately to the epoch of decoupling and its value will approximate to this epoch ($\approx 3.77\times10^5$~yrs) with an upper limit at the age of the oldest red-shift galaxies thus far discovered, such as GLASS-z12 that is redshifted to $z=12.4^{+0.1}_{-0.3}$ and had already built up $\sim10^9 M\odot$ in stars $\lesssim 300-400$ Myr after the Big Bang \citep{2022ApJ...940L..14N}, or the age of M92 at 13~Gyr \cite{2023arXiv230104659W}. 
At $T_g$, local expansion of proto-galactic clounds as a consequence of the expansion of the Universe stopped and galactic separation began, and it may be asserted that the proto-galaxy then had a mean stellar motion of zero relative to the centre of the proto-galaxy.
The present day separation distance between galaxies is influenced by the continual expansion of space between them after individual galaxies separated, leading to the familiar Hubble's Law of red-shift with distance, and by mutual gravitational attraction between neighbouring galaxies with the infalling galactic clusters leading to galactic interactions or ultimately to complex galactic mergers. 

Taking the present age of the universe $T_0\sim13.8$~Gyr \cite{2018arXiv180706209P} and the mean radius of the proto-galaxies as $r_0$, then the mean separation of the proto-galaxies at $T_g$ was $2r_0$.
Let the mean separation between galactic centres be $d_g$ at the present day, and taking an upper limit to the epoch of galaxy formation ($T_g$) as $\sim6\times 10^8$~yrs, then $d_g=2r_0 T_0/T_g\sim 46 r_0$ through Hubble separation. 
Substituting $r_f\sim1.14 r_0$ (from equation~\ref{eq:epsilon}) suggests a present ratio of mean galaxy separation distance to mean galactic radius of $d_g/r_f=46/(2\times1.14)\sim20$. 

\section{The Jeans Equation and Faber-Jackson relationship}
The first proposition of the Jeans theorem states: 
\textit{Any steady-state solution of the collisionless Boltzmann equation depends on the phase-space coordinates only through integrals of motion in the galactic potential. Any function of the integrals yields a steady-state solution of the collisionless Botlzmann equation} \cite{2008gady.book.....B}.
The collisionless state is justified for GCs because collisions between their stars are rare, as confirmed by the very low rate of super novae \cite{2012A&A...539A..77V}.

Since we rarely know more than three integrals, this definition is often modified to the second proposition of Jeans's theorem: 
\textit{Any function of integrals solves the collisionless Boltzmann equation}~\cite{2008gady.book.....B}: 
\begin{equation}
    \rho \frac{\partial \overline{v}_j}{\partial t}
+\rho \overline{v}_i\frac{\partial \overline{v}_j}{\partial x_i}
=-\rho\frac{\partial \Phi}{\partial x_j}-\frac{\partial}{\partial x_i}(\rho \sigma^2_{ij})
\label{eq:Jeans}
\end{equation}
where position $x=(x_1, x_2, x_3)$, $v_{i/j}=(v_1, v_2, v_3)$ is the velocity, $\Phi=\Phi(x,t)$ is the gravitational potential, $\rho =\rho_{x,t}$ is the local density at time $t$, and $\sigma_{i,j}$ is the velocity dispersion tensor, modified from Eq.~4-27 of \cite{2008gady.book.....B}.

We have derived a function for total mass rather than luminosity, and from equation~\ref{eq:v_bar} we may estimate the total mass $M_0$ as:
\begin{equation}
	M_0=\frac{\pi{}\,r_f\overline{v}^2}{2G}\,.
	\label{eq:M_0}
\end{equation}

Equation \ref{eq:M_0} is similar to the spherical Jeans equation in one dimension assuming spherical hydrostatic equilibrium and isotropy.
The velocity dispersion tensor, $\sigma_{ij}$, is then:  $\sigma_{rr}^2=\sigma{_{t1}^2}=\sigma_0^2$.
But the galaxy may not be spherically symmetrical and may also be anisotropic with radial velocities as well as tangential ones. 
In addition, spherical galaxies, especially fainter ones, don't possess clearly defined boundaries making the final radius, $r_f$, difficult to measure.

Wolf {\it{et al}} (2010) show that the integrated mass within a characteristic radius, $r_3$, where the log-slope of the 3D density profile is $-3$, is largely insensitive to velocity dispersion anisotropy and for a wide range of stellar light distributions that describe dispersion-supported galaxies, $r_3$ is close to the 3D de-projected half-light radius $r_{1/2}$ \cite{2010MNRAS.406.1220W}. 
We therefore define $r_f$ as a multiple of the effective half-light radius ($R_{1/2}$), or the radius containing half the galaxy's luminosity. 
Because the graph of $M$--$R_{1/2}$ is plotted on a log-log scale, the precise value of $\beta$ is less critical and a commonly accepted approximation $\beta\simeq2$ is used.
This may be related by 
\begin{equation}
   r_f=\beta R_{1/2} \textnormal{,  where }1 < \beta \lesssim2\,. \label{eq:Rf}
\end{equation}

Elliptical galaxies are generally smooth, featureless systems containing Population II stars and little or no gas or dust, giving them some similarities to GCs. 
They are denoted by their degree of eccentricity from the most spherical, $E0$, to $E7$, the most elongated, but because we see only their brightness distribution it is impossible to tell directly if they are axisymmetric or triaxial \cite{2008gady.book.....B}.

\subsection{Spherical and elliptical galaxies}
The Faber-Jackson (F-J) relation provides a power-law relation between luminosity $L$ and the central stellar velocity dispersion $\sigma_0$ of elliptical galaxies. 
Their relation can be expressed mathematically as: 
\begin{equation}
    L\propto \sigma_0^\gamma \textnormal{~~~with~~}\gamma\approx4\,.
    \label{eq:FJ}
\end{equation}

\begin{figure}
\centering
    \includegraphics[width=0.5\columnwidth]{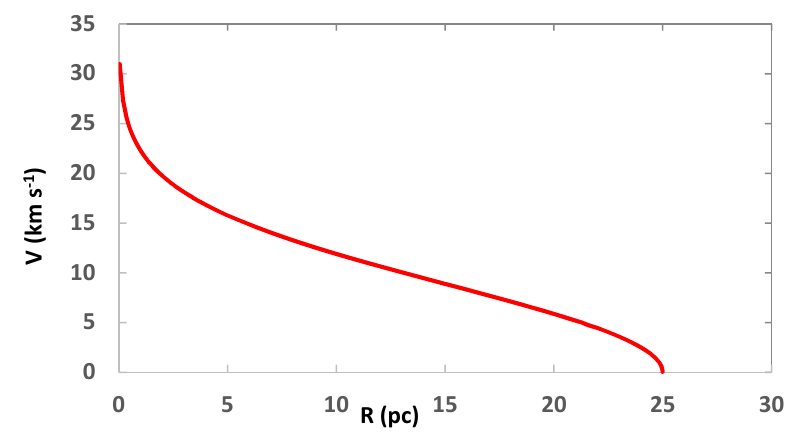}
    \vspace{0cm}
    \caption{Velocity-Radius curve for M15}
    \label{fig:VR}
\end{figure}
\begin{figure}
\centering
    \includegraphics [width=0.5\columnwidth]{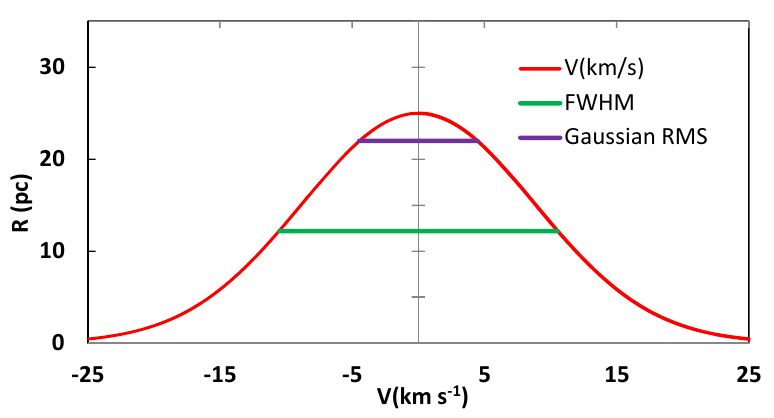}
    \vspace{0cm}
    \caption{Fig. \ref{fig:VR} as a Gaussian curve across the whole GC. }
    \label{fig:Gaussian}
\end{figure}
\begin{figure*}
\centering
    \includegraphics[width=\textwidth]{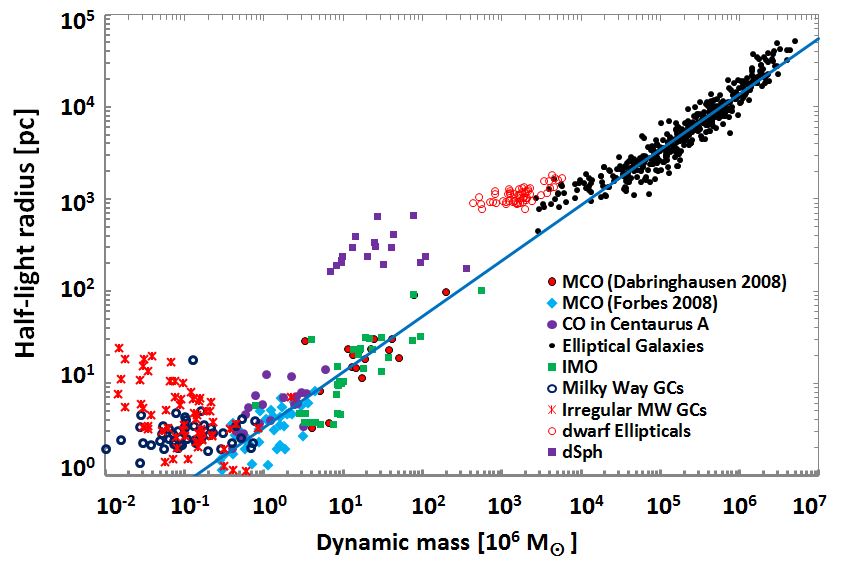}
    \vspace{0cm}
    \caption{Log-log plot of half-light radius/Mass for a range of spherical and elliptical galaxies. 
    The radius containing half of each object's light, $R-{1/2}$, is plotted against each object's stellar mass. 
    Blue open circles: MW Globular Clusters; purple circles: Compact Objects (CO) in Centaurus A; black circles: dwarf ellipticals (dE) and ordinary ellipticals; green squares: Intermediate Mass Objects (IMO); blue diamonds: Massive Compact Objects (MCO); red circles: MCO; red crosses: Irregular or low mass MW GCs; purple squares: dwarf spherical galaxies (dSph). 
    Blue line is linear best fit to ordinary ellipticals and compact objects (\textit{see text for references}).
    }
    \label{fig:ellipticals}
\end{figure*}

The F-J relation linking the mass and luminosity of an elliptical galaxy to its velocity dispersion specifically refers to the rms width of the Gaussian distribution function of the line-of-sight velocity from a selected range of absorption bands \cite{1976ApJ...204..668F}. In contrast to $\sigma_0$ in the F-B relation, $\overline{v}$ (equation~\ref{eq:v_bar}) is a mean velocity across the whole system, and better refers to the Full Width at Half Maximum (FWHM) value. 
The velocity/radius curve for a typical star in M15 is shown in Fig.~\ref{fig:VR}.
By switching axes, Fig.~\ref{fig:Gaussian} shows the velocity distribution across the whole M15 galaxy which approximates to the expected Gaussian distribution.
Fig.~\ref{fig:Gaussian} also shows the rms and FWHM values which are related by:
\begin{equation}
    \overline{v}=2\sqrt{2\ln2}\sigma_0\,.
\end{equation}

The total dynamic mass $M_0$ may therefore be estimated from equation~\ref{eq:M_0} as:
\begin{equation}
    M_0\simeq \frac{4\pi \ln2 \sigma_0^2 R_{1/2}}{G}\,,
    \label{eq:M0}
\end{equation}
where $\sigma_0$ is the rms line-of-sight velocity dispersion of the galaxy, and $R_{1/2}$ is the half-light radius.

The parameters $\sigma_0$ and $R_{1/2}$ are listed in many catalogues for a wide range of galactic masses and types.
Fig.~\ref{fig:ellipticals} plots the derived log-log dynamic mass vs. half-light radius for 735 spherical objects spanning more than 8 decades of mass.
Data for these plots and the number in each class, [n], were taken for: dwarf ellipticals (dE) [58] and ordinary ellipticals [437] \cite{2008MNRAS.389.1924F}; Massive Compact Objects (MCO) [55] \cite{2008MNRAS.389.1924F, 2008MNRAS.386..864D}; Compact Objects (CO) [20] in Centaurus A \cite{2008MNRAS.386..864D}; Intermediate Mass Objects (IMO) [32] \cite{2008MNRAS.389.1924F}; MW GCs [48] \cite{2008MNRAS.389.1924F}; Irregular, low mass or very faint MW GCs [67] \cite{2018MNRAS.478.1520B}; and dwarf spherical galaxies (dSph) [18] \cite{1993AJ....105..510M, 2008MNRAS.389.1924F}.
\begin{figure*}
	\includegraphics[width=\textwidth]{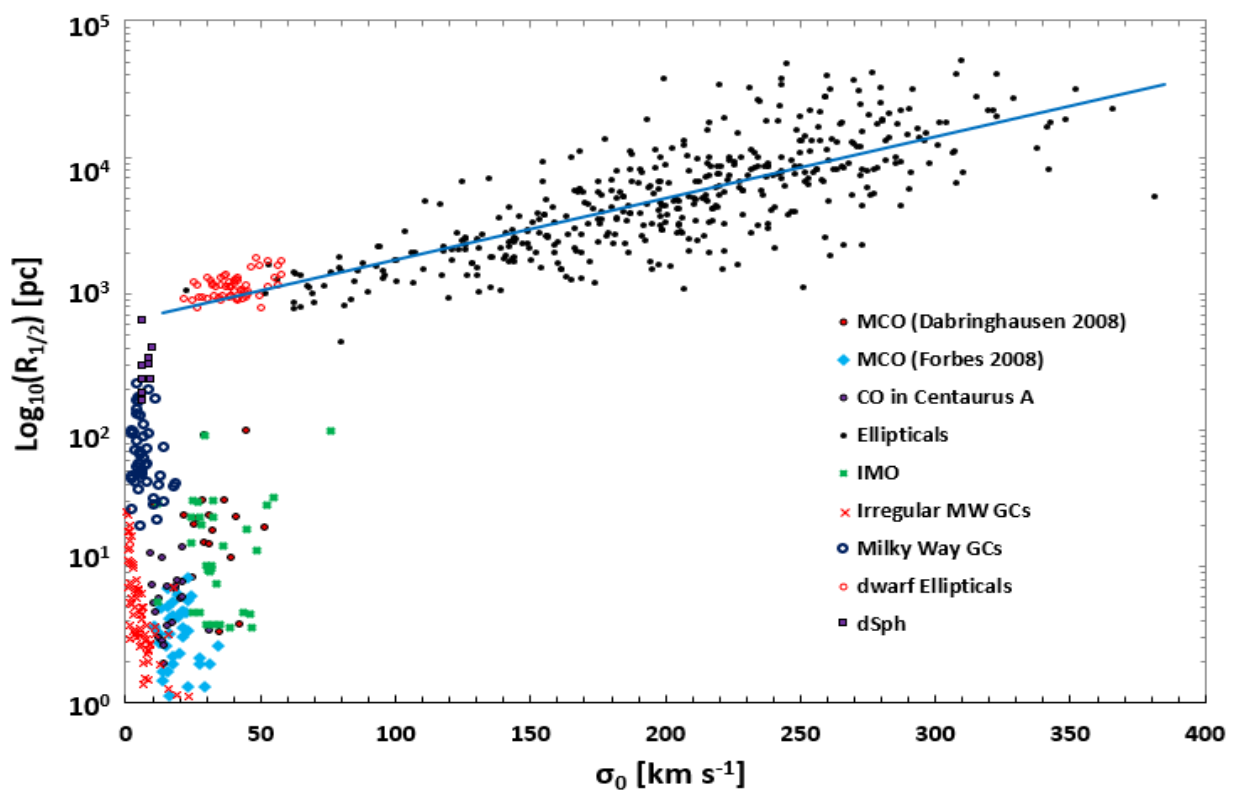}
    \vspace{0cm}
    \caption{Semi-Log plot of $R_{1/2}$--$\sigma_0$ for galaxies of Fig.~\ref{fig:ellipticals} with rms best fit (blue solid line) for the ellipticals and dwarf ellipticals (\textit{Abbreviations as Fig.~\ref{fig:ellipticals}. For references see text}). 
    }
    \label{fig:R_sigma}
\end{figure*}

Models assuming structural homology, such as de Vaucouleurs' $R^{1/4}$ model, cannot establish changing trends in their slope \cite{2013pss6.book...91G}.
An F-J slope $\sim4$ represents the average slope over a restricted luminosity range to what is a curved or broken $L-\sigma^\gamma$ distribution, with the slope becoming $\sim2$ at lower luminosities. 
Also shown in Fig.~\ref{fig:ellipticals} is the linear best fit to the normal ellipticals and compact objects (excluding dwarf ellipticals, dSphs and GCs). 
This has a mean $R-M_{vir}$ log-log slope of ${0.604\pm0.003}$, equivalent to an F-J slope of $\gamma=3.66{\pm}0.01$ over a range of 7 decades.
Despite the good fit to a majority of the galaxies in Fig.~\ref{fig:ellipticals}, there are some noted anomalies with the dwarf ellipticals, dSphs and GCs branching from the main line. 

\subsection{Anomolous groups}
The difference in scaling relations between elliptical galaxies and globular clusters demonstrated in Figs.~\ref{fig:ellipticals} and \ref{fig:R_sigma} is well-known and documented \cite{1995ApJ...438L..29D, 2005ApJ...627..203H}. 
Dwarf elliptical galaxies (dE) have blue absolute magnitudes within the range $-18$ mag $< M_b < -14$ mag, fainter than ordinary elliptical galaxies. 
Objects with masses intermediate between massive GCs and dwarf ellipticals have masses of $\sim10^7$~M$_\odot$ and relatively compact sizes. They are usually referred to as Ultra Compact Dwarfs (UCDs), with sizes intermediate between elliptical galaxies and globular clusters.
\cite{2008MNRAS.389.1924F} argue that most of these intermediate mass objects have similar properties to massive GCs, i.e. IMOs are essentially massive star clusters.
The dSphs have luminosities of order $10^5-10^7$~L$_\odot$, and are characterised by their low surface brightnesses.

In galaxy formation theory based on the cold dark matter (CDM) model, low-luminosity, low-mass galaxies such as dSphs are considered to be among the first bound luminous objects and are therefore expected to contain DM, and current data for dSph galaxies are consistent with these systems having similar DM halos, with total masses in the range $M_{vir}<10^8$~M$_{\odot}$ \cite{1993AJ....105..510M}.
A "discontinuity" is seen at $M_{vir}=10^8$~M$_\odot$.
This has been interpreted as the threshold for the gas in dSphs to be blown away by successive supernovae \cite{Hirashita_1998}.
The large mass ratio of DM to baryonic matter is therefore reached in a low mass dSph through gas depletion.

In contrast to the F-J diagram of Fig.~\ref{fig:ellipticals}, Fig.~\ref{fig:R_sigma} is a semi-log plot of $\log R_{1/2}$--$\sigma_o$.
The elliptical and dwarf ellipticals galaxies now form a slowly increasing continuum over approximately two decades of $R_{1/2}$ as $\sigma_0$ increases over a range of $20\lesssim\sigma_0\gtrsim380$ km sec$^{-1}$.
This is in marked distinction from the other groups of galaxies, where $R_{1/2}$ increases by nearly three decades while $\sigma_0$ remains low, never being $\gtrsim 70$ km sec$^{-1}$.
This contrasting pattern is consistent with the ellipticals containing a black hole core whose mass increases as the velocity dispersion increases, compared with the remaining types of spherical or irregular galaxies which have an absent or low-mass core.

Metz and Kroupa (2007) have shown that modelling tidal dwarf galaxies (TDGs) results in objects that resemble known dSphs after a Hubble time of dynamical evolution \cite{2007MNRAS.376..387M}. 
During each close galactic passage, energy is pumped into the dSph leading to an expansion of the satellite, leaving a quasi-stable remnant that matches many of the dSphs of the MW and Andromeda.
Such mechanisms may also account for the irregular Milky Way GCs apparent in Fig.~\ref{fig:R_sigma}.

Dwarf ellipticals present another anomaly on the $\log R-\log M$ graph, with a gradual break away from the trend line at $M_{vir}\approx 5\times10^9$~M$_\odot$.
In contrast, their good fit to the rms trend line for ellipticals in  Fig.~\ref{fig:R_sigma} suggests they are a continuation of the ellipticals, and reexpansion after mass loss (either by a supernova-driven wind or by ram-pressure stripping) may account for their anisotropic velocity dispersions. 
Alternately, the proto-systems from which dwarf ellipticals formed might not have isotropized their velocity dispersions due to very low initial density and, in consequence, a low collision rate within the molecular cloud system \cite{1992ApJ...399..462B}.

\section{Discussion}
This paper presents an idealised  model for a typical GC to reach stability, from the time of separation from the expanding Universe in an initially uniform state until its formation as an independent spherical entity.
The velocity/radius relationship is plotted, and the resulting log-log surface-density/radius curves compare well with observations for three typical GCs.
Given the velocity dispersion and half-light radius, the model enables the dynamical mass to be computed, and the model was extended to plot this theoretical mass against the half-light radius for 735 spherical objects spanning more than eight decades of mass (Fig.~\ref{fig:ellipticals}).

The absolute age of the Universe is a cosmological limit setting a maximum age to all objects within it.
The epoch of decoupling (379,000 years after the Big Bang) sets a further age limit because, apart from the small ripples recently  observed in the MWB, the whole Universe was then at a state of equilibrium \cite{2018arXiv180706209P}. 
Although this is uncomfortably close to the ages of some observed components of the Universe such as some of the oldest GCs, some supermassive black hole (SMBH) quasars, and some of the oldest stars (Section~\ref{Chronology}), leaving a very short time for them to have condensed from the expanding hot gas field, there is increasing theoretical evidence that this can occur \cite{2020MNRAS.498.5652K}.
Current theory states that the initial separation occurred from the gaseous state after decoupling, but the determined ages of objects such as primordial stars, and the short time available for the collapse of early SMBHs, are consistent with separation beginning within the plasma itself before the era of decoupling, and such differentiation, even at the size of a SMBH, would still be too small to register on the large scale primordial ripples of the CMB.
Gravitational instability then acted upon minor perturbations within this medium allowed the separation of individual pro-galactic units that went on to collapse through their own self gravity against the background of an expanding universe. 
This caused the mean separation of galactic centres to continue to increase even as the galaxies themselves contracted in size. 

Whatever the mechanism for the formation of individual galaxies, it is generally accepted that they must have condensed out of a medium that was, at least initially, approximately homogeneous, and that all massive galaxies host a SMBH at their centres.
There is also increasing evidence that these SMBHs co-evolved with
their host galaxies, with an empirical scaling relation between the SMBH mass and galaxy properties such as the stellar velocity dispersion \cite{2020MNRAS.494.1784A}, while disk galaxies also appear to have a universal spin parameter \cite{2008gady.book.....B, 2015MNRAS.453.2214M}.
However, if disc galaxies obtained their angular momentum early in their evolution, this may have been greater than they could sustain.
They could then attain a stable spin by removing both mass and angular momentum from the primary disc, leading to the ejection of a number of distinct and separate orbiting satellite galaxies, including GCs and dSphs, implying that these GCs are coequal in age with their massive parent proto-disc galaxies.

Modelling GCs suggests that all spherically-symmetrical assemblies in gravitational equilibrium and without angular momentum attained a stable, final maximum radius that is a universal constant, $\simeq 1.136$ times their initial radius at the epoch of galactic separation.
Their predicted density and velocity distribution match observational data reasonably well, and it is demonstrated that the ratio of mean separation distance:galactic radius is a constant for a given epoch, with a plausible present day value $d_g/r_g\sim20$.

A number of theoretical predictions suggest the existence of DM to make up missing mass both from the universe and from within galaxies, including DM within spherical galaxies, where the mass/velocity dispersion relationship with DM is similar to that plotted in Fig.~\ref{fig:ellipticals} \cite{2022IJAA...12..258H}.
The model presented in this paper does not preclude the existence of DM in GCs, but it does suggest that the distribution of any DM follows the distribution of the observed stellar masses.
It is not the purpose of this paper to consider missing DM candidates or their detailed behaviour; however, it follows that any DM must be cold-DM particles that are gravitationally bound within the GC or galaxy. 
Within a GC, such particles will behave similarly to the 'particles' of the star systems, i.e. they will exchange potential for kinetic energy as they move towards the centre under gravity, and they will gravitationally react with other particles to exchange momentum in an elastic, non-dissipative manner to reach an overall thermal equilibrium. 
With the congregation of more massive objects towards the centre, the DM may be confined to a thin, undetected DM halo which would increase the effective radius of the GC.
Modelling GCs, however, does suggest that they do not require DM halos nor any MOND-like modification to Newtonian gravitation.

\vspace{6pt} 



The author declares no conflict of interest. 

\bibliographystyle{mdpi}
\bibliography{GCs}

\begin{thebibliography}{999}

\bibitem[{Gaia Collaboration} et~al.(2021){Gaia Collaboration}, {Brown}, {Vallenari}, {Prusti}, and {et~al}]{2021A&A...649A...1G}
{Gaia Collaboration}.; {Brown}, A.G.A.; {Vallenari}, A.; {Prusti}, T.; {et~al}.
\newblock {Gaia Early Data Release 3. Summary of the contents and survey properties}.
\newblock {\em \aap} {\bf 2021}, {\em 649},~A1,  \href{http://arxiv.org/abs/2012.01533}{{\normalfont [arXiv:astro-ph.GA/2012.01533]}}.
\newblock {\url{https://doi.org/10.1051/0004-6361/202039657}}.

\bibitem[{Piatti} et~al.(2023){Piatti}, {Webb}, and {Carlberg}]{2023yCat..74894367P}
{Piatti}, A.E.; {Webb}, J.J.; {Carlberg}, R.G.
\newblock {VizieR Online Data Catalog: Radii of the Milky Way globular clusters (Piatti+, 2019)}.
\newblock {\em VizieR Online Data Catalog} {\bf 2023}, p. J/MNRAS/489/4367.

\bibitem[{Chandrasekhar}(1942)]{1942idlb.book..313C}
{Chandrasekhar}, J., {Principles of Stellar Dynamics}.
\newblock In {\em Principles of Stellar Dynamics}, 1st ed.; Dover Publications, New York, N.Y,  1942; p. 313.

\bibitem[{Smail}(2020)]{2020...Smail}
{Smail}, I.
\newblock The {Hertzsprung-Russell} Diagram of a Globular Cluster.
\newblock http://community.dur.ac.uk/ian.smail/gcCm/gcCm\_summ.html,  2020.
\newblock Accessed: 2024-01-22.

\bibitem[{Chaisson} and {McMillan}(1999)]{1999idlb.book..832C}
{Chaisson}, E.J.; {McMillan}, S., {Astronomy Today}.
\newblock In {\em Astronomy Today}, 3rd ed.; Prentice Hall,  1999; p. 832.

\bibitem[{Weisz} et~al.(2023){Weisz}, {McQuinn}, {Savino}, and {et~al}]{2023arXiv230104659W}
{Weisz}, D.R.; {McQuinn}, K.B.W.; {Savino}, A.; {et~al}.
\newblock {The JWST Resolved Stellar Populations Early Release Science Program II. Survey Overview}.
\newblock {\em arXiv e-prints} {\bf 2023}, p. arXiv:2301.04659,  \href{http://arxiv.org/abs/2301.04659}{{\normalfont [arXiv:astro-ph.GA/2301.04659]}}.
\newblock {\url{https://doi.org/10.48550/arXiv.2301.04659}}.

\bibitem[{von Hippel} et~al.(2001){von Hippel}, {Simpson}, and {Manset}]{2001ASPC..245..162C}
{von Hippel}, T.; {Simpson}, C.; {Manset}, N., Eds.
\newblock {\em Globular Cluster Age Dating}, Vol. 245, {\em Astronomical Society of the Pacific Conference Series},  2001.

\bibitem[{Deiters} et~al.(2001){Deiters}, {Fuchs}, {Just}, {Spurzem}, and {Wielen}]{2001ASPC..228.....D}
{Deiters}, S.; {Fuchs}, B.; {Just}, A.; {Spurzem}, R.; {Wielen}, R.
\newblock Multi-mass Gaseous Models of Globular Clusters with Stellar Evolution.
\newblock In Proceedings of the Dynamics of Star Clusters and the Milky Way. ASPC Conference Series 228,  2001, Vol. 228.

\bibitem[{Zonoozi} et~al.(2014){Zonoozi}, {Haghi}, {K{\"u}pper}, {Baumgardt}, {Frank}, and {Kroupa}]{2014MNRAS.440.3172Z}
{Zonoozi}, A.H.; {Haghi}, H.; {K{\"u}pper}, A.H.W.; {Baumgardt}, H.; {Frank}, M.J.; {Kroupa}, P.
\newblock {Direct N-body simulations of globular clusters - II. Palomar 4}.
\newblock {\em \mnras} {\bf 2014}, {\em 440},~3172--3183,  \href{http://arxiv.org/abs/1404.5969}{{\normalfont [arXiv:astro-ph.GA/1404.5969]}}.
\newblock {\url{https://doi.org/10.1093/mnras/stu526}}.

\bibitem[{Binney} and {Tremaine}(2008)]{2008gady.book.....B}
{Binney}, J.; {Tremaine}, S.
\newblock {\em Galactic Dynamics: Second Edition}; Princeton U. Press, Princeton N.J.,  2008.

\bibitem[{Hut}(1997)]{1997Comp....3.1H}
{Hut}, P.
\newblock {Gravitational Thermodynamics}.
\newblock {\em Complexity} {\bf 1997}, {\em 3},~1.

\bibitem[{Planck Collaboration} et~al.(2020){Planck Collaboration}, {Aghanim}, {Akrami}, {Ashdown}, {Aumont}, {Baccigalupi}, {Ballardini}, {Banday}, and {et. al.}]{2020A&A...641A...6P}
{Planck Collaboration}.; {Aghanim}, N.; {Akrami}, Y.; {Ashdown}, M.; {Aumont}, J.; {Baccigalupi}, C.; {Ballardini}, M.; {Banday}, A.J.; {et. al.}.
\newblock {Planck 2018 results. VI. Cosmological parameters}.
\newblock {\em \aap} {\bf 2020}, {\em 641},~A6,  \href{http://arxiv.org/abs/1807.06209}{{\normalfont [arXiv:astro-ph.CO/1807.06209]}}.
\newblock {\url{https://doi.org/10.1051/0004-6361/201833910}}.

\bibitem[{Bennett} et~al.(2013){Bennett}, {Larson}, {Weiland}, {Jarosik}, {Hinshaw}, {Odegard}, and {et. al}]{2013ApJS..208...20B}
{Bennett}, C.L.; {Larson}, D.; {Weiland}, J.L.; {Jarosik}, N.; {Hinshaw}, G.; {Odegard}, N.; {et. al}.
\newblock {Nine-year Wilkinson Microwave Anisotropy Probe (WMAP) Observations: Final Maps and Results}.
\newblock {\em \apjs} {\bf 2013}, {\em 208},~20,  \href{http://arxiv.org/abs/1212.5225}{{\normalfont [arXiv:astro-ph.CO/1212.5225]}}.
\newblock {\url{https://doi.org/10.1088/0067-0049/208/2/20}}.

\bibitem[{Hinshaw} et~al.(2009){Hinshaw}, {Weiland}, {Hill}, {Odegard}, and {et~al}]{2009ApJS..180..225H}
{Hinshaw}, G.; {Weiland}, J.L.; {Hill}, R.S.; {Odegard}, N.; {et~al}.
\newblock {Five-Year Wilkinson Microwave Anisotropy Probe Observations: Data Processing, Sky Maps, and Basic Results}.
\newblock {\em \apjs} {\bf 2009}, {\em 180},~225--245,  \href{http://arxiv.org/abs/0803.0732}{{\normalfont [arXiv:astro-ph/0803.0732]}}.
\newblock {\url{https://doi.org/10.1088/0067-0049/180/2/225}}.

\bibitem[{Planck Collaboration}(2016)]{2016A&A...594A..13P}
{Planck Collaboration}.
\newblock {Planck 2015 results. XIII. Cosmological parameters}.
\newblock {\em \aap} {\bf 2016}, {\em 594},~A13,  \href{http://arxiv.org/abs/1502.01589}{{\normalfont [arXiv:astro-ph.CO/1502.01589]}}.
\newblock {\url{https://doi.org/10.1051/0004-6361/201525830}}.

\bibitem[{Planck Collaboration}(2018)]{2018arXiv180706209P}
{Planck Collaboration}.
\newblock {Planck 2018 results. VI. Cosmological parameters}.
\newblock {\em arXiv e-prints} {\bf 2018}, p. arXiv:1807.06209,  \href{http://arxiv.org/abs/1807.06209}{{\normalfont [arXiv:astro-ph.CO/1807.06209]}}.

\bibitem[{Finkelstein} et~al.(2022){Finkelstein}, {Bagley}, {Ferguson}, {Wilkins}, and {et~al}]{2022arXiv221105792F}
{Finkelstein}, S.L.; {Bagley}, M.B.; {Ferguson}, H.C.; {Wilkins}, S.M.; {et~al}.
\newblock {CEERS Key Paper I: An Early Look into the First 500 Myr of Galaxy Formation with JWST}.
\newblock {\em arXiv e-prints} {\bf 2022}, p. arXiv:2211.05792,  \href{http://arxiv.org/abs/2211.05792}{{\normalfont [arXiv:astro-ph.GA/2211.05792]}}.
\newblock {\url{https://doi.org/10.48550/arXiv.2211.05792}}.

\bibitem[{Oesch} et~al.(2016){Oesch}, {Brammer}, {van Dokkum}, {Illingworth}, and {et~al}]{2016ApJ...819..129O}
{Oesch}, P.A.; {Brammer}, G.; {van Dokkum}, P.G.; {Illingworth}, G.D.; {et~al}.
\newblock {A Remarkably Luminous Galaxy at z=11.1 Measured with Hubble Space Telescope Grism Spectroscopy}.
\newblock {\em \apj} {\bf 2016}, {\em 819},~129,  \href{http://arxiv.org/abs/1603.00461}{{\normalfont [arXiv:astro-ph.GA/1603.00461]}}.
\newblock {\url{https://doi.org/10.3847/0004-637X/819/2/129}}.

\bibitem[{Wang} et~al.(2021){Wang}, {Yang}, {Fan}, and {et~al}]{2021ApJ...907L...1W}
{Wang}, F.; {Yang}, J.; {Fan}, X.; {et~al}.
\newblock {A Luminous Quasar at Redshift 7.642}.
\newblock {\em \apjl} {\bf 2021}, {\em 907},~L1,  \href{http://arxiv.org/abs/2101.03179}{{\normalfont [arXiv:astro-ph.GA/2101.03179]}}.
\newblock {\url{https://doi.org/10.3847/2041-8213/abd8c6}}.

\bibitem[{Woods} et~al.(2019){Woods}, {Agarwal}, {Bromm}, {Bunker}, and {et~al}]{2019PASA...36...27W}
{Woods}, T.E.; {Agarwal}, B.; {Bromm}, V.; {Bunker}, A.; {et~al}.
\newblock {Titans of the early Universe: The Prato statement on the origin of the first supermassive black holes}.
\newblock {\em \pasa} {\bf 2019}, {\em 36},~e027,  \href{http://arxiv.org/abs/1810.12310}{{\normalfont [arXiv:astro-ph.GA/1810.12310]}}.
\newblock {\url{https://doi.org/10.1017/pasa.2019.14}}.

\bibitem[{Kroupa} et~al.(2020){Kroupa}, {Subr}, {Jerabkova}, and {Wang}]{2020MNRAS.498.5652K}
{Kroupa}, P.; {Subr}, L.; {Jerabkova}, T.; {Wang}, L.
\newblock {Very high redshift quasars and the rapid emergence of supermassive black holes}.
\newblock {\em \mnras} {\bf 2020}, {\em 498},~5652--5683,  \href{http://arxiv.org/abs/2007.14402}{{\normalfont [arXiv:astro-ph.GA/2007.14402]}}.
\newblock {\url{https://doi.org/10.1093/mnras/staa2276}}.

\bibitem[Tang and Joyce(2021)]{Tang_2021}
Tang, J.; Joyce, M.
\newblock Revised Best Estimates for the Age and Mass of the Methuselah Star HD 140283 Using MESA and Interferometry and Implications for 1D Convection.
\newblock {\em Research Notes of the AAS} {\bf 2021}, {\em 5},~117.
\newblock {\url{https://doi.org/10.3847/2515-5172/ac01ca}}.

\bibitem[{Valcin} et~al.(2020){Valcin}, {Bernal}, {Jimenez}, {Verde}, and {Wandelt}]{2020JCAP...12..002V}
{Valcin}, D.; {Bernal}, J.L.; {Jimenez}, R.; {Verde}, L.; {Wandelt}, B.D.
\newblock {Inferring the age of the universe with globular clusters}.
\newblock {\em \jcap} {\bf 2020}, {\em 2020},~002,  \href{http://arxiv.org/abs/2007.06594}{{\normalfont [arXiv:astro-ph.CO/2007.06594]}}.
\newblock {\url{https://doi.org/10.1088/1475-7516/2020/12/002}}.

\bibitem[{Alexander} and {Natarajan}(2014)]{2014Sci...345.1330A}
{Alexander}, T.; {Natarajan}, P.
\newblock {Rapid growth of seed black holes in the early universe by supra-exponential accretion}.
\newblock {\em Science} {\bf 2014}, {\em 345},~1330--1333,  \href{http://arxiv.org/abs/1408.1718}{{\normalfont [arXiv:astro-ph.GA/1408.1718]}}.
\newblock {\url{https://doi.org/10.1126/science.1251053}}.

\bibitem[Gerssen et~al.(2002)Gerssen, van~der Marel, Gebhardt, Guhathakurta, Peterson, and Pryor]{Gerssen_2002}
Gerssen, J.; van~der Marel, R.P.; Gebhardt, K.; Guhathakurta, P.; Peterson, R.C.; Pryor, C.
\newblock {Hubble Space Telescope Evidence for an Intermediate-Mass Black Hole in the Globular Cluster M15. {II}. Kinematic Analysis and Dynamical Modeling}.
\newblock {\em \aj} {\bf 2002}, {\em 124},~3270--3288.
\newblock {\url{https://doi.org/10.1086/344584}}.

\bibitem[Giesers et~al.(2018)Giesers, Dreizler, Husser, and {et~al}]{10.1093/mnrasl/slx203}
Giesers, B.; Dreizler, S.; Husser, T.O.; {et~al}.
\newblock {A detached stellar-mass black hole candidate in the globular cluster NGC 3201}.
\newblock {\em Monthly Notices of the Royal Astronomical Society: Letters} {\bf 2018}, {\em 475},~L15--L19,  \href{http://arxiv.org/abs/https://academic.oup.com/mnrasl/article-pdf/475/1/L15/24841598/slx203.pdf}{{\normalfont [https://academic.oup.com/mnrasl/article-pdf/475/1/L15/24841598/slx203.pdf]}}.
\newblock {\url{https://doi.org/10.1093/mnrasl/slx203}}.

\bibitem[{Voss} and {Nelemans}(2012)]{2012A&A...539A..77V}
{Voss}, R.; {Nelemans}, G.
\newblock {Type Ia supernovae in globular clusters: observational upper limits}.
\newblock {\em \aap} {\bf 2012}, {\em 539},~A77,  \href{http://arxiv.org/abs/1111.6593}{{\normalfont [arXiv:astro-ph.HE/1111.6593]}}.
\newblock {\url{https://doi.org/10.1051/0004-6361/201118222}}.

\bibitem[{Marr}(2015)]{2015MNRAS.448.3229M}
{Marr}, J.H.
\newblock {Galaxy rotation curves with lognormal density distribution}.
\newblock {\em \mnras} {\bf 2015}, {\em 448},~3229--3241,  \href{http://arxiv.org/abs/1502.02949}{{\normalfont [arXiv:astro-ph.GA/1502.02949]}}.
\newblock {\url{https://doi.org/10.1093/mnras/stv216}}.

\bibitem[McNamara et~al.(2004)McNamara, Harrison, and Baumgardt]{McNamara_2004}
McNamara, B.J.; Harrison, T.E.; Baumgardt, H.
\newblock {The Dynamical Distance to M15: Estimates of the Cluster's Age and Mass and of the Absolute Magnitude of Its {RR} Lyrae Stars}.
\newblock {\em The Astrophysical Journal} {\bf 2004}, {\em 602},~264--270.
\newblock {\url{https://doi.org/10.1086/380905}}.

\bibitem[{Baratpour} and {Khodadadi}(2012)]{2012JStatRes...B}
{Baratpour}, S.; {Khodadadi}, A.
\newblock {A Cumulative Residual Entropy Characterization of the Rayleigh Distribution and Related Goodness-of-Fit Test}.
\newblock {\em J. Statist. Res. Iran} {\bf 2012}, {\em 9},~115--131.

\bibitem[{Salpeter}(1955)]{1955ApJ...121..161S}
{Salpeter}, E.E.
\newblock {The Luminosity Function and Stellar Evolution.}
\newblock {\em \apj} {\bf 1955}, {\em 121},~161.
\newblock {\url{https://doi.org/10.1086/145971}}.

\bibitem[{Chabrier}(2005)]{2005ASSL..327...41C}
{Chabrier}, G., {The Initial Mass Function: From Salpeter 1955 to 2005}.
\newblock In {\em The Initial Mass Function 50 Years Later}; Astrophysics and Space Science Library,  2005; Vol. 327, p.~41.
\newblock {\url{https://doi.org/10.1007/978-1-4020-3407-7_5}}.

\bibitem[{Bolzonella} et~al.(2000){Bolzonella}, {Miralles}, and {Pell{\'o}}]{2000A&A...363..476B}
{Bolzonella}, M.; {Miralles}, J.M.; {Pell{\'o}}, R.
\newblock {Photometric redshifts based on standard SED fitting procedures}.
\newblock {\em \aap} {\bf 2000}, {\em 363},~476--492,  \href{http://arxiv.org/abs/astro-ph/0003380}{{\normalfont [arXiv:astro-ph/astro-ph/0003380]}}.

\bibitem[{Chabrier}(2003)]{2003PASP..115..763C}
{Chabrier}, G.
\newblock {Galactic Stellar and Substellar Initial Mass Function}.
\newblock {\em \pasp} {\bf 2003}, {\em 115},~763--795,  \href{http://arxiv.org/abs/astro-ph/0304382}{{\normalfont [arXiv:astro-ph/astro-ph/0304382]}}.
\newblock {\url{https://doi.org/10.1086/376392}}.

\bibitem[{Kroupa} et~al.(2013){Kroupa}, {Weidner}, {Pflamm-Altenburg}, {Thies}, {Dabringhausen}, {Marks}, and {Maschberger}]{2013pss5.book..115K}
{Kroupa}, P.; {Weidner}, C.; {Pflamm-Altenburg}, J.; {Thies}, I.; {Dabringhausen}, J.; {Marks}, M.; {Maschberger}, T.
\newblock {The Stellar and Sub-Stellar Initial Mass Function of Simple and Composite Populations}. In {\em Planets, Stars and Stellar Systems. Volume 5: Galactic Structure and Stellar Populations}; {Oswalt}, T.D.; {Gilmore}, G., Eds.; Springer Refernce,  2013; Vol.~5, p. 115.
\newblock {\url{https://doi.org/10.1007/978-94-007-5612-0_4}}.

\bibitem[{Miller} and {Scalo}(1979)]{1979ApJS...41..513M}
{Miller}, G.E.; {Scalo}, J.M.
\newblock {The Initial Mass Function and Stellar Birthrate in the Solar Neighborhood}.
\newblock {\em \apjs} {\bf 1979}, {\em 41},~513.
\newblock {\url{https://doi.org/10.1086/190629}}.

\bibitem[{King}(1962)]{1962AJ.....67..471K}
{King}, I.
\newblock {The structure of star clusters. I. an empirical density law}.
\newblock {\em \aj} {\bf 1962}, {\em 67},~471.
\newblock {\url{https://doi.org/10.1086/108756}}.

\bibitem[{Djorgovski} and {King}(1984)]{1984ApJ...277L..49D}
{Djorgovski}, S.; {King}, I.R.
\newblock {Surface photometry in cores of globular clusters.}
\newblock {\em \apjl} {\bf 1984}, {\em 277},~L49--L52.
\newblock {\url{https://doi.org/10.1086/184200}}.

\bibitem[{Newell} et~al.(1976){Newell}, {Da Costa}, and {Norris}]{1976ApJ...208L..55N}
{Newell}, B.; {Da Costa}, G.S.; {Norris}, J.
\newblock {Evidence for a Central Massive Object in the X-Ray Cluster M15}.
\newblock {\em \apjl} {\bf 1976}, {\em 208},~L55.
\newblock {\url{https://doi.org/10.1086/182232}}.

\bibitem[{Di~Cecco} et~al.(2013){Di~Cecco}, Zocchi, Varri, Monelli, and {et~al}]{Di_Cecco_2013}
{Di~Cecco}, A.; Zocchi, A.; Varri, A.L.; Monelli, M.; {et~al}.
\newblock On The Density Profile Of The Globular Cluster M92*.
\newblock {\em The Astronomical Journal} {\bf 2013}, {\em 145},~103.
\newblock {\url{https://doi.org/10.1088/0004-6256/145/4/103}}.

\bibitem[{Naidu} et~al.(2022){Naidu}, {Oesch}, {van Dokkum}, {Nelson}, {Suess}, {Brammer}, and {et. al.Whitaker}]{2022ApJ...940L..14N}
{Naidu}, R.P.; {Oesch}, P.A.; {van Dokkum}, P.; {Nelson}, E.J.; {Suess}, K.A.; {Brammer}, G.; {et. al.Whitaker}.
\newblock {Two Remarkably Luminous Galaxy Candidates at z {\ensuremath{\approx}} 10-12 Revealed by JWST}.
\newblock {\em \apjl} {\bf 2022}, {\em 940},~L14,  \href{http://arxiv.org/abs/2207.09434}{{\normalfont [arXiv:astro-ph.GA/2207.09434]}}.
\newblock {\url{https://doi.org/10.3847/2041-8213/ac9b22}}.

\bibitem[{Wolf} et~al.(2010){Wolf}, {Martinez}, {Bullock}, {Kaplinghat}, {Geha}, {Mu{\~n}oz}, {Simon}, and {Avedo}]{2010MNRAS.406.1220W}
{Wolf}, J.; {Martinez}, G.D.; {Bullock}, J.S.; {Kaplinghat}, M.; {Geha}, M.; {Mu{\~n}oz}, R.R.; {Simon}, J.D.; {Avedo}, F.F.
\newblock {Accurate masses for dispersion-supported galaxies}.
\newblock {\em \mnras} {\bf 2010}, {\em 406},~1220--1237,  \href{http://arxiv.org/abs/0908.2995}{{\normalfont [arXiv:astro-ph.CO/0908.2995]}}.
\newblock {\url{https://doi.org/10.1111/j.1365-2966.2010.16753.x}}.

\bibitem[{Faber} and {Jackson}(1976)]{1976ApJ...204..668F}
{Faber}, S.M.; {Jackson}, R.E.
\newblock {Velocity dispersions and mass-to-light ratios for elliptical galaxies.}
\newblock {\em \apj} {\bf 1976}, {\em 204},~668--683.
\newblock {\url{https://doi.org/10.1086/154215}}.

\bibitem[{Forbes} et~al.(2008){Forbes}, {Lasky}, {Graham}, and {Spitler}]{2008MNRAS.389.1924F}
{Forbes}, D.A.; {Lasky}, P.; {Graham}, A.W.; {Spitler}, L.
\newblock {Uniting old stellar systems: from globular clusters to giant ellipticals}.
\newblock {\em \mnras} {\bf 2008}, {\em 389},~1924--1936,  \href{http://arxiv.org/abs/0806.1090}{{\normalfont [arXiv:astro-ph/0806.1090]}}.
\newblock {\url{https://doi.org/10.1111/j.1365-2966.2008.13739.x}}.

\bibitem[{Dabringhausen} et~al.(2008){Dabringhausen}, {Hilker}, and {Kroupa}]{2008MNRAS.386..864D}
{Dabringhausen}, J.; {Hilker}, M.; {Kroupa}, P.
\newblock {From star clusters to dwarf galaxies: the properties of dynamically hot stellar systems}.
\newblock {\em \mnras} {\bf 2008}, {\em 386},~864--886,  \href{http://arxiv.org/abs/0802.0703}{{\normalfont [arXiv:astro-ph/0802.0703]}}.
\newblock {\url{https://doi.org/10.1111/j.1365-2966.2008.13065.x}}.

\bibitem[{Baumgardt} and {Hilker}(2018)]{2018MNRAS.478.1520B}
{Baumgardt}, H.; {Hilker}, M.
\newblock {A catalogue of masses, structural parameters, and velocity dispersion profiles of 112 Milky Way globular clusters}.
\newblock {\em \mnras} {\bf 2018}, {\em 478},~1520--1557,  \href{http://arxiv.org/abs/1804.08359}{{\normalfont [arXiv:astro-ph.GA/1804.08359]}}.
\newblock {\url{https://doi.org/10.1093/mnras/sty1057}}.

\bibitem[{Mateo} et~al.(1993){Mateo}, {Olszewski}, {Pryor}, {Welch}, and {Fischer}]{1993AJ....105..510M}
{Mateo}, M.; {Olszewski}, E.W.; {Pryor}, C.; {Welch}, D.L.; {Fischer}, P.
\newblock {The Carina Dwarf Spheroidal Galaxy: How Dark is it?}
\newblock {\em \aj} {\bf 1993}, {\em 105},~510.
\newblock {\url{https://doi.org/10.1086/116449}}.

\bibitem[{Graham}(2013)]{2013pss6.book...91G}
{Graham}, A.W., {Elliptical and Disk Galaxy Structure and Modern Scaling Laws}.
\newblock In {\em Planets, Stars and Stellar Systems. Volume 6: Extragalactic Astronomy and Cosmology}; Springer Netherlands,  2013; Vol.~6, p.~91.
\newblock {\url{https://doi.org/10.1007/978-94-007-5609-0_2}}.

\bibitem[{Djorgovski}(1995)]{1995ApJ...438L..29D}
{Djorgovski}, S.
\newblock {The Fundamental Plane Correlations for Globular Clusters}.
\newblock {\em \apjl} {\bf 1995}, {\em 438},~L29.
\newblock {\url{https://doi.org/10.1086/187707}}.

\bibitem[{Ha{\c{s}}egan} et~al.(2005){Ha{\c{s}}egan}, {Jord{\'a}n}, {C{\^o}t{\'e}}, {Djorgovski}, {McLaughlin}, {Blakeslee}, {Mei}, {West}, {Peng}, {Ferrarese}, {Milosavljevi{\'c}}, {Tonry}, and {Merritt}]{2005ApJ...627..203H}
{Ha{\c{s}}egan}, M.; {Jord{\'a}n}, A.; {C{\^o}t{\'e}}, P.; {Djorgovski}, S.G.; {McLaughlin}, D.E.; {Blakeslee}, J.P.; {Mei}, S.; {West}, M.J.; {Peng}, E.W.; {Ferrarese}, L.;  et~al.
\newblock {The ACS Virgo Cluster Survey. VII. Resolving the Connection between Globular Clusters and Ultracompact Dwarf Galaxies}.
\newblock {\em \apj} {\bf 2005}, {\em 627},~203--223,  \href{http://arxiv.org/abs/astro-ph/0503566}{{\normalfont [arXiv:astro-ph/astro-ph/0503566]}}.
\newblock {\url{https://doi.org/10.1086/430342}}.

\bibitem[Hirashita et~al.(1998)Hirashita, Takeuchi, and Tamura]{Hirashita_1998}
Hirashita, H.; Takeuchi, T.T.; Tamura, N.
\newblock Physical Interpretation of the Mass-Luminosity Relation of Dwarf Spheroidal Galaxies.
\newblock {\em The Astrophysical Journal} {\bf 1998}, {\em 504},~L83--L86.
\newblock {\url{https://doi.org/10.1086/311584}}.

\bibitem[{Metz} and {Kroupa}(2007)]{2007MNRAS.376..387M}
{Metz}, M.; {Kroupa}, P.
\newblock {Dwarf spheroidal satellites: are they of tidal origin?}
\newblock {\em \mnras} {\bf 2007}, {\em 376},~387--392,  \href{http://arxiv.org/abs/astro-ph/0701289}{{\normalfont [arXiv:astro-ph/astro-ph/0701289]}}.
\newblock {\url{https://doi.org/10.1111/j.1365-2966.2007.11438.x}}.

\bibitem[{Bender} et~al.(1992){Bender}, {Burstein}, and {Faber}]{1992ApJ...399..462B}
{Bender}, R.; {Burstein}, D.; {Faber}, S.M.
\newblock {Dynamically Hot Galaxies. I. Structural Properties}.
\newblock {\em \apj} {\bf 1992}, {\em 399},~462.
\newblock {\url{https://doi.org/10.1086/171940}}.

\bibitem[{Asmus} et~al.(2020){Asmus}, {Greenwell}, {Gandhi}, {Boorman}, and {et al}]{2020MNRAS.494.1784A}
{Asmus}, D.; {Greenwell}, C.L.; {Gandhi}, P.; {Boorman}, P.G.; {et al}.
\newblock {Local AGN survey (LASr): I. Galaxy sample, infrared colour selection, and predictions for AGN within 100 Mpc}.
\newblock {\em \mnras} {\bf 2020}, {\em 494},~1784--1816,  \href{http://arxiv.org/abs/2003.05959}{{\normalfont [arXiv:astro-ph.GA/2003.05959]}}.
\newblock {\url{https://doi.org/10.1093/mnras/staa766}}.

\bibitem[{Marr}(2015)]{2015MNRAS.453.2214M}
{Marr}, J.H.
\newblock {Angular momentum of disc galaxies with a lognormal density distribution}.
\newblock {\em \mnras} {\bf 2015}, {\em 453},~2214--2219,  \href{http://arxiv.org/abs/1507.04515}{{\normalfont [arXiv:astro-ph.GA/1507.04515]}}.
\newblock {\url{https://doi.org/10.1093/mnras/stv1734}}.

\bibitem[{Hoeneisen}(2022)]{2022IJAA...12..258H}
{Hoeneisen}, B.
\newblock {Measurement of the Dark Matter Velocity Dispersion with Galaxy Stellar Masses, UV Luminosities, and Reionization}.
\newblock {\em International Journal of Astronomy and Astrophysics} {\bf 2022}, {\em 12},~258--272,  \href{http://arxiv.org/abs/2209.03168}{{\normalfont [arXiv:physics.gen-ph/2209.03168]}}.
\newblock {\url{https://doi.org/10.4236/ijaa.2022.123015}}.

\end{thebibliography}
\end{document}